\documentclass[12pt]{article}

\usepackage{graphics,epsfig,subfigure}
\usepackage{amssymb,amsmath,bm}

\def\simle{\mathrel{\rlap{\raise 0.511ex \hbox{$<$}}{\lower 0.511ex
 \hbox{$\sim$}}}}

\newcommand{\bea}{\begin{eqnarray}}
\newcommand{\eea}{\end{eqnarray}}
\newcommand{\be}{\begin{equation}}
\newcommand{\ee}{\end{equation}}
\newcommand{\nn}{\nonumber}
\newcommand{\gev}{{\rm~GeV}}
\newcommand{\mev}{{\rm~MeV}}
\newcommand{\fm}{{\rm~fm}}
\newcommand{\msb}{\overline{\rm{MS}}}
\newcommand{\mbar}{\overline{m}}

\def\rmii{a}
\def\rmiiinfn{b}
\def\mad{c}
\def\rmiii{d}
\def\infn{e}
\def\liv{f}
\def\orsay{g}
\def\hum{h}

\begin{document}

\begin{titlepage}
{
\vspace{-0.5cm}
\normalsize
\hfill \parbox{105mm}{ROM2F/2011/08, FTUAM-11-51, IFT-UAM/CSIC-11-53,\\ RM3-TH/11-4, HU-EP-11/29, SFB/CPP-11-35}}\\[10mm]
\vspace{-0.5cm}
\begin{center}
  \begin{Large}
\boldmath    
{\bf Lattice QCD determination of $m_b$, $f_B$ and $f_{Bs}$ \\
            with twisted mass Wilson fermions\\}
\unboldmath
  \end{Large}
\end{center}

\vspace{-0.5cm}
\begin{figure}[h]
  \begin{center}
    \includegraphics[draft=false]{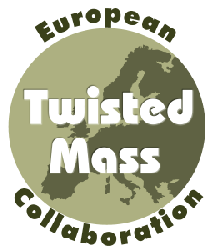}
  \end{center}
\end{figure}

\vspace{-0.5cm}
\baselineskip 20pt plus 2pt minus 2pt
\begin{center}
  \textbf{
    P.~Dimopoulos$^{(\rmii,\rmiiinfn)}$,
    R.~Frezzotti$^{(\rmii,\rmiiinfn)}$,
    G.~Herdoiza$^{(\mad)}$,
    V.~Lubicz$^{(\rmiii,\infn)}$,\\
    C.~Michael$^{(\liv)}$,
    D.~Palao$^{(\rmiiinfn)}$,
    G.~C.~Rossi$^{(\rmii,\rmiiinfn)}$,
    F.~Sanfilippo$^{(\orsay)}$,\\
    A.~Shindler$^{(\hum)}$\footnote{Heisenberg Fellow},
    S.~Simula$^{(\infn)}$,
    C.~Tarantino$^{(\rmiii,\infn)}$,
    M.~Wagner$^{(\hum)}$}\\
\end{center}

\begin{center}
  \begin{footnotesize}
    \noindent

$^{(\rmii)}$ Dip. di Fisica, Universit{\`a} di Roma Tor Vergata, Via della Ricerca Scientifica, I-00133 Roma, Italy
\vspace{0.2cm}

$^{(\rmiiinfn)}$ INFN, Sez. di Roma Tor Vergata, Via della Ricerca Scientifica, I-00133 Roma, Italy
\vspace{0.2cm}

$^{(\mad)}$ Departamento de F\'isica Te\'orica and Instituto de F\'isica Te\'orica UAM/CSIC,\\
Universidad Aut\'onoma de Madrid, Cantoblanco, E-28049 Madrid, Spain 
\vspace{0.2cm}

$^{(\rmiii)}$ Dip. di Fisica, Universit{\`a} Roma Tre, Via della Vasca Navale
84, I-00146 Roma, Italy
\vspace{0.2cm}

$^{(\infn)}$ INFN, Sez. di Roma
Tre, Via della Vasca Navale 84, I-00146 Roma, Italy
\vspace{0.2cm}

$^{(\liv)}$ Theoretical Physics Division, Dept. of Mathematical Sciences,\\
University of Liverpool, Liverpool L69 7ZL, United Kingdom
\vspace{0.2cm}

$^{(\orsay)}$ Laboratoire de Physique Th\'eorique (Bat. 210), Universit\'e Paris Sud,\\ Centre d'Orsay, F-91405 Orsay-Cedex, France
\vspace{0.2cm}

$^{(\hum)}$ Humboldt Universit\"at zu Berlin, Newtonstrasse 15, D-12489, Berlin, Germany
\vspace{0.2cm}

  \end{footnotesize}
\end{center}

\begin{abstract}
We present a lattice QCD determination of the $b$ quark mass and of the $B$ and $B_s$ decay constants, performed with $N_f=2$ twisted mass Wilson fermions, by simulating at four values of the lattice spacing. In order to study the $b$ quark on the lattice, two methods are adopted in the present work, respectively based on suitable ratios with exactly known static limit and on the interpolation between relativistic data, evaluated in the charm mass region, and the static point, obtained by simulating the HQET on the lattice. The two methods provide results in good agreement. For the $b$ quark mass in the $\msb$ scheme and for the decay constants we obtain $\mbar_b(\mbar_b)=4.29(14)\,\gev$, $f_B=195(12)\,\mev$, $f_{Bs}=232(10)\,\mev$ and $f_{Bs}/f_B=1.19(5)$. As a byproduct of the analysis we also obtain the results for the $f_D$ and $f_{Ds}$ decay constants: $f_D=212(8)\,\mev$, $f_{Ds}=248(6)\,\mev$ and $f_{Ds}/{f_D}=1.17(5)$.

\end{abstract}
\end{titlepage}

\section{Introduction}
The study of physical processes involving the $b$ quark are of utmost importance for accurate tests of  the Standard Model and for searching New Physics effects. On the experimental side, B-factories have played a fundamental role in the achievement of the present accuracy and further improvements are expected and looked forward from LHCb and the planned SuperB factories. It is therefore crucial to have theoretical uncertainties well under control, in particular those of the hadronic parameters computed on the lattice.

Two particularly important cases of study are the purely leptonic decays $B \to \tau \nu_\tau$ and $B_s \to \mu^+ \mu^-$. The first process is particularly sensitive to potential New Physics contributions mediated, at tree level, by charged Higgs. The relevant entries in the Standard Model prediction for the decay rate are the CKM matrix element $V_{ub}$, which can be extracted from the study of semileptonic $B\to \pi\, \ell\, \nu_{\ell}$ decays without significant New Physics contributions (for $\ell=e,\mu$), and the pseudoscalar decay constant $f_B$. The measured values of the $B \to \tau \nu_\tau$ decay rate deviate, at present, by about 3 sigma from the Standard Model prediction~\cite{utfit,Bona:2009cj}, within relatively large experimental and theoretical uncertainties. In this respect, improving the lattice determination of $f_B$ would be an important ingredient for increasing the chances of detecting the contribution of New Physics effects to this decay. Another golden process for the detection of potentially large New Physics contributions is the rare leptonic decay $B_s \to \mu^+ \mu^-$, which is being studied with unprecedented accuracy at LHCb. In this case, the relevant hadronic parameter to be determined on the lattice, which enters the theoretical prediction of the decay rate, is the pseudoscalar decay constant $f_{B_s}$. The determination of both $f_B$ and $f_{B_s}$, together with a prediction for the $b$ quark mass $m_b$, are the scope of the present study.

With the available computer power it is not possible to simulate quark masses in the range of the physical $b$ mass keeping, at the same time, finite volume and discretization effects under control. In order to circumvent these problems, many different methods have been proposed so far (see~\cite{Aubin:2009yh} for a recent review). 

In~\cite{Blossier:2009hg} we performed an exploratory calculation of the $b$ quark mass and the decay constants $f_B$ and $f_{Bs}$ by introducing suitable ratios having an exactly known static limit. In~\cite{Blossier:2009gd} a more standard method~\cite{Allton:1990qg} was applied, using lattice QCD data with the heavy quark mass ranging from the charm region up to more than $3\,\gev$, together with the information coming from a calculation in the static limit point. In the following, we will refer to the two approaches as to the  ``ratio method" and the ``interpolation method" respectively.

Here we update and finalize both the analyses, by implementing several improvements. We replace the preliminary values of the quark mass renormalization constants with the published results of~\cite{Constantinou:2010gr}. We increase for some ensembles the statistics  and we use more data, in particular, data at the finest lattice spacing ($\beta=4.2$) are now included also in the analysis with the ratio method. The main improvement in the analysis based on the interpolation method consists in studying the dependence of the decay constants on the quark masses, instead of the meson masses, and performing the extrapolation to the continuum limit at fixed (reference) values of the heavy quark mass. This allows us to better disentangle discretization effects from the (physical) heavy quark mass dependence. The use of the quark masses in the determination of $f_B$ and $f_{Bs}$ requires as input the value of the $b$ quark mass, $m_b$, which we obtain from the ratio method.

Our results for the $b$ quark mass in the $\msb$ scheme and for the decay constants (the latter obtained by averaging the results of the two methods) read
\bea
&\mbar_b(\mbar_b)=4.29(14)\,\gev\,,&\nn\\ 
&&\nn\\
&f_B=195(12)\,\mev\,,\quad f_{Bs}=232(10)\,\mev\,,\quad \dfrac{f_{Bs}}{f_B}=1.19(5)\,,&
\eea
which are in good agreement with our previous results~\cite{Blossier:2009hg,Blossier:2009gd}, but which have smaller uncertainties.
In particular, with respect to the result $\mbar_b(\mbar_b)=4.63(27)\,\gev$ of~\cite{Blossier:2009hg}, we obtain for the $b$ quark mass a central value which is smaller by approximately one standard deviation and a reduction of the uncertainty by almost a factor two, mainly because of the improvement in the determination of the quark mass renormalization constant.

As a byproduct of the analysis we also obtain the results for the $f_D$ and $f_{Ds}$ decay constants
\be
f_D=212(8)\,\mev\,, \qquad \qquad f_{Ds}=248(6)\,\mev\,, \qquad \qquad \frac{f_{Ds}}{f_D}=1.17(5)\,,
\ee
which update and improve our previous determination~\cite{Blossier:2009bx}.

\section{Simulation details}
The calculation is based on the $N_f=2$ gauge field configurations generated by the European Twisted Mass (ETM) Collaboration with the tree-level improved Symanzik gauge action~\cite{Weisz82}  and the twisted mass quark action~\cite{FGSW01}  at maximal twist, discussed in detail in~\cite{Baron:2009wt}-\cite{Frezzotti:2003ni}. We simulated $N_f=2$ mass-degenerate dynamical quarks, whose mass is eventually extrapolated to the physical isospin averaged mass of the up and down quarks, $m_{u/d}$. The strange and charm quarks are quenched in the present calculation.
In our lattice setup all physical quantities are $\mathcal{O}(a)$ improved~\cite{Frezzotti:2003ni}, in particular cutoff effects related to the heavy quark mass $\mu_h$ are of order $a^2 \mu_h^2$. 

For further details of our simulations we refer to~\cite{Blossier:2010cr}, where the same ensembles of gauge configurations were used. We recall here that data at four values of the lattice coupling, $\beta=\{3.80, 3.90, 4.05, 4.20\}$, are included in the analysis. The corresponding values of the lattice spacing, $a=\{0.098(3), 0.085(2), 0.067(2), 0.054(1)\}\,\fm$, have been determined in~\cite{Blossier:2010cr} together with $m_{u/d}$ using the physical values of the pion mass and decay constant as input. From~\cite{Blossier:2010cr} we also take the values of the average up/down and the strange quark masses, namely $\mbar_{u/d}(2 \gev)=3.6(2)\,\mev$ and $\mbar_s(2 \gev)=95(6)\,\mev$. For the quark mass renormalization constants $Z_{\mu}=Z_P^{-1}$ we use the results obtained in~\cite{Constantinou:2010gr}, i.e. $Z_P(\msb, 2\gev)=\{0.411(12), 0.437(7), 0.477(6), 0.501(20)\}$ at the four beta values (see also~\cite{Blossier:2010cr} for the estimate of $Z_P$ at $\beta=4.20$).

At variance with~\cite{Blossier:2010cr}, where only the light, strange and charm quark masses were studied, a wider range of values for the valence quark masses is considered here, in order to get closer to the physical $b$ quark mass. The values of the simulated valence quark masses are collected in Table~\ref{tab:val}.
The values of the valence light quark mass, $\mu_\ell$, are always taken identical to those of their sea counterparts.
 The heavy quark mass $\mu_h$ ranges from approximately $m_c$ up to $2.3-2.4\,m_c$, being $m_c$ the physical charm quark mass. Correlators at higher $\mu_h$ values have been simulated and were included in~\cite{Blossier:2009gd}. They are characterized by large fluctuations in the effective mass plateaux, and thus by large statistical uncertainties. As a consequence, these data turn out to be irrelevant in the fits, and we have excluded them from the present analysis.
%%%%%%%%%%%%%%%%%%%%%%%%%%%%%%%%%%%%%%%%%%%%%%%%%%%%%%%%%%%
\begin{table}[!t]
\begin{center}
\renewcommand{\arraystretch}{1.4}
\begin{tabular}{||c||c|c|c|c||}
\hline
$\beta$  &  $a \mu_\ell$ &   $a \mu_s$    & $a \mu_h$ & $t_{min}/a$\\ \hline\hline
3.80 & 0.0080, 0.0110 & 0.0165, 0.0200& 0.2143, 0.2406, 0.2701, 0.3032 & 14 \\
        &                          &     0.0250         & 0.3403,  0.3819, 0.4287, 0.4812 & \\ \hline
3.90 & 0.0030, 0.0040,& 0.0150, 0.0180 & 0.2049, 0.2300, 0.2582, 0.2898 & 16 \\
        &  0.0064, 0.0085, 0.0100 &  0.0220 & 0.3253,  0.3651, 0.4098, 0.4600 & \\ \hline
4.05 & 0.0030, 0.0060, & 0.0135, 0.0150,& 0.1663, 0.1867, 0.2096, 0.2352 & 21 \\
        &  0.0080             & 0.0180            &  0.2640, 0.2963, 0.3326, 0.3733 & \\ \hline
4.20 & 0.0020, 0.0065 & 0.0130, 0.0148   & 0.1477, 0.1699, 0.1954, 0.2247 & 25 \\
        &                          &0.0180               & 0.2584, 0.2971, 0.3417               & \\ \hline
\hline
\end{tabular}
\renewcommand{\arraystretch}{1.0}
\end{center}
\vspace{-0.4cm}
\caption{\sl Values of simulated bare quark masses in lattice units, for the four $\beta$ values, in the light ($a \mu_\ell$), strange  ($a \mu_s$) and heavy ($a \mu_h$) sectors. In the last column the minimum values of time $t_{min}$ chosen for the 2-point function fits are collected.}
\label{tab:val}
\end{table}
%%%%%%%%%%%%%%%%%%%%%%%%%%%%%%%%%%%%%%%%%%%%%%%%%%%%%%%%%%%

We now proceed to describe the two approaches adopted in the present work to study the $B$-physics observables, namely the ratio method of~\cite{Blossier:2009hg} and the interpolation method.

\section{Ratio method}
\subsection{The $\mathbf{b}$ quark mass}
The $b$ quark mass is obtained by implementing the ratio method of~\cite{Blossier:2009hg}, briefly recalled hereafter. The method is suggested by the HQET asymptotic behavior of the heavy-light meson mass $M_{h\ell}$,
\be
\lim_{\mu_h^{\rm pole}\to \infty} \left( \frac{M_{h\ell}}{\mu_h^{\rm pole}} \right)={\mbox{constant}} \ ,
\label{PM}
\ee
where $\mu_h^{\rm pole}$ is the pole quark mass and the limit~(\ref{PM}) is approached without corrections of $\mathcal{O}(1/\log(\mu_h^{\rm pole}/\Lambda_{\rm QCD}))$.
The first step is to consider an appropriate sequence of heavy quark masses, $\bar\mu_h^{(1)},\, \bar\mu_h^{(2)},\, \ldots \,,\, \bar\mu_h^{(N)}$, with fixed ratio
\be
\frac{\bar\mu_h^{(n)}}{\bar\mu_h^{(n-1)}} = \lambda \ ,
\label{eq:hqmasses}
\ee
and ranging from the charm mass to values somewhat below the bottom mass.
Here and in the following we denote by a ``bar'' the quark masses renormalized in the $\overline{MS}$ scheme and, if not otherwise specified, a renormalization scale of $2\,\gev$ is implied. 

Then one computes the following ratios that have an exactly known static limit,
\bea
y(\bar\mu_h^{(n)},\lambda;\bar\mu_{\ell},a) & \equiv & 
\frac{M_{h\ell}(\bar\mu_h^{(n)};\bar\mu_\ell,a)}{M_{h\ell}(\bar\mu_h^{(n-1)};\bar\mu_\ell,a)} \cdot 
\frac{\bar\mu^{(n-1)}_h}{\bar\mu^{(n)}_h} \cdot
\frac{\rho( \bar\mu^{(n-1)}_h,\mu^*)}{\rho( \bar\mu^{(n)}_h,\mu^*)}=\nn\\
&=& \lambda^{-1} \frac{M_{h\ell}(\bar\mu_h^{(n)};\bar\mu_\ell,a)}{M_{h\ell}(\bar\mu^{(n)}_h/\lambda;\bar\mu_\ell,a)} 
\cdot \frac{ \rho( \bar\mu^{(n)}_h/\lambda,\mu^* )}{\rho( \bar\mu^{(n)}_h,\mu^*)}\, ,\quad\quad n=2,\cdots,N \, .
\label{RATM}
\eea
The function $\rho(\bar\mu_h,\mu^*)$ is the factor that converts the renormalized $\overline{MS}$ quark mass (at the scale $\mu^*$) into the pole mass, 
\be
\mu_h^{\rm{pole}} = \rho(\bar\mu_h, \mu^*) \, \bar\mu_h(\mu^*)\ ,
\label{PMA}
\ee
known up to N$^3$LO in perturbation theory~\cite{CR,POLEtoMSbarMASS}.
The NLO expression reads
\bea
&&\rho(\bar\mu_h, \mu^*)=\Biggl[1+\frac{16}{3}\cdot\frac{\alpha^{\overline{\mbox{MS}}}(\bar \mu_h)}{4\pi}\Biggr]\cdot\Biggl(\frac{\alpha^{\overline{\mbox{MS}}}(\bar \mu_h)}{\alpha^{\overline{\mbox{MS}}}( \mu^*)}\Biggr)^{12/(33-2\,N_f)}\,\cdot\nn\\
&&
\Biggl[1+\Biggl(\frac{2(4491-252\,N_f+20\,N_f^2)}{3(33-2\,N_f)^2} \Biggr)\, \frac{\alpha^{\overline{\mbox{MS}}}(\bar \mu_h)-\alpha^{\overline{\mbox{MS}}}(\mu^*)}{4\pi}\Biggr]\,,
\label{eq:rho}
\eea
used for  $N_f=2$ in the present analysis.
We notice that the dependence on the scale $\mu^*$ cancels in the ratios of $\rho$ factors evaluated at different heavy quark masses and thus in the $y$ ratio defined in eq.~(\ref{RATM}).

From eq.~(\ref{PM}) and QCD asymptotic freedom it follows that the ratios~(\ref{RATM}) have the following static limit:
\be
\lim_{\bar\mu_h \to \infty} y(\bar\mu_h,\lambda;\bar\mu_{\ell},a=0) =1 \,.
\label{PM2}
\ee

The value of the ratio $\lambda$ of eq.~(\ref{eq:hqmasses}), between two subsequent values of the heavy quark mass, is chosen in such a way that after a finite number of steps the heavy-light meson mass assumes the experimental value $M_B=5.279\gev$ (we find $\lambda=1.1762$) . In order to implement this condition, the lattice data at the four lattice spacings are interpolated at the following values of the heavy quark mass,
\bea
&&\bar\mu_h^{(1)}= 1.140~{\mbox{GeV}} \  , \
\bar\mu_h^{(2)}= \lambda\, \bar\mu_h^{(1)} = 1.341~{\mbox{GeV}} \  , \
\bar\mu_h^{(3)}= \lambda^2 \bar\mu_h^{(1)} = 1.577~{\mbox{GeV}}\, ,\\
&&\bar\mu_h^{(4)}= \lambda^3 \bar\mu_h^{(1)} = 1.855~{\mbox{GeV}} \  , \
\bar\mu_h^{(5)}= \lambda^4 \bar\mu_h^{(1)} = 2.182~{\mbox{GeV}}  \  , \
\bar\mu_h^{(6)}= \lambda^5 \bar\mu_h^{(1)} = 2.566~{\mbox{GeV}}\, . \nn
\label{MUV}\eea

\begin{figure}[t]	
\includegraphics[width=0.48\textwidth]{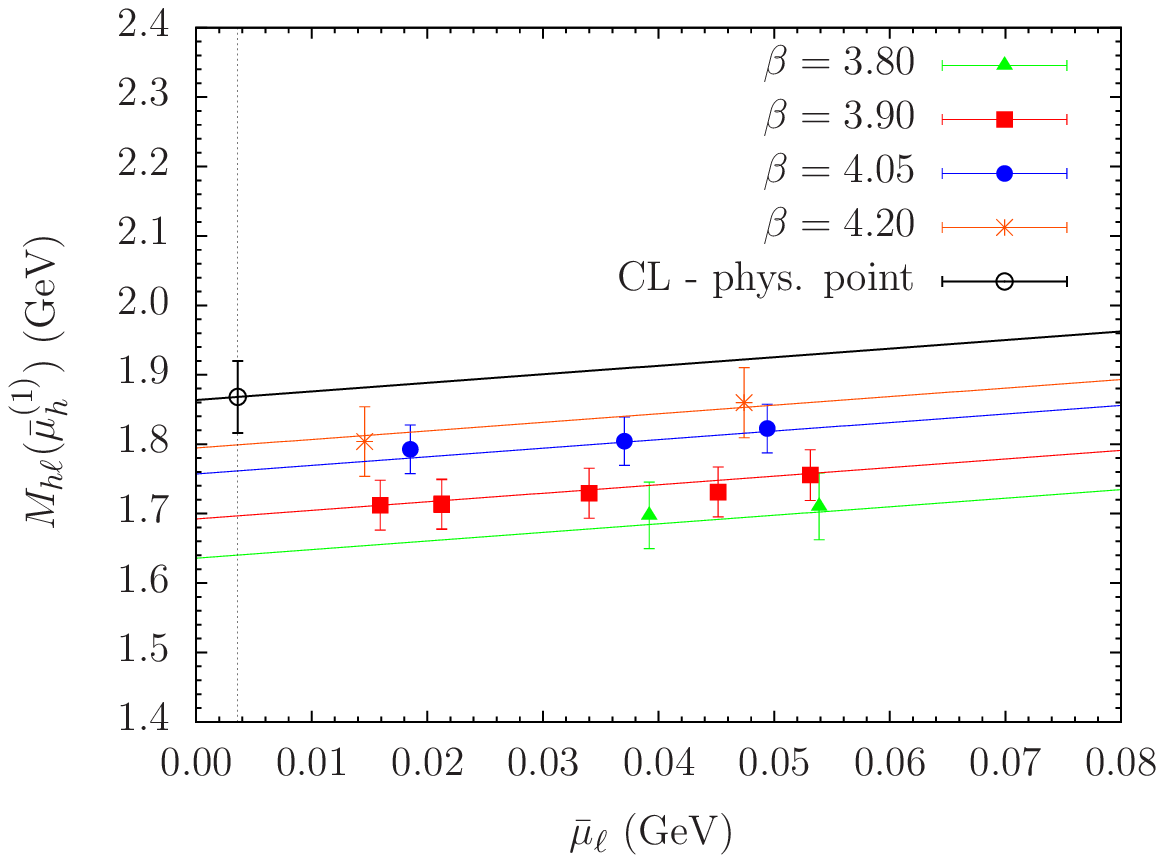}
\includegraphics[width=0.48\textwidth]{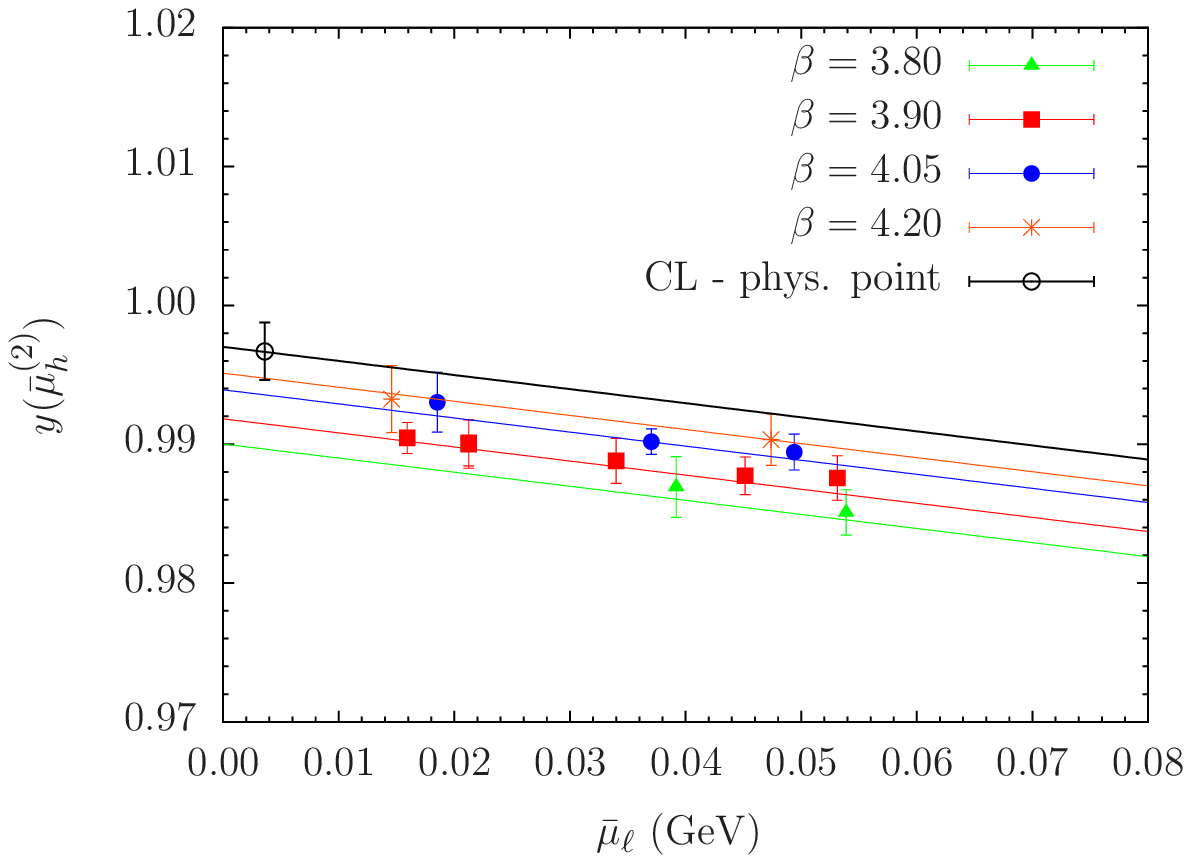}
\caption{\sl Light quark mass dependence of the meson mass $M_{h\ell}(\bar\mu_h^{(1)})$ (left) and of the ratio $y(\bar\mu_h^{(2)})$ (right) at the four values of the lattice spacing.}
\label{fig:Mmul}
\end{figure}

Ratios of the kind defined in eq.~(\ref{RATM}) are introduced because, besides having an exactly known static limit, they are also expected~\cite{Blossier:2009hg} to have a smooth chiral and continuum limit, as shown in the right plot of fig.~\ref{fig:Mmul}. 
In fig.~\ref{fig:Mmul} (left) we show the chiral and continuum extrapolation of the heavy-light meson mass evaluated at the first of our heavy quark masses $\bar \mu_h^{(1)}$, namely $M_{hl}(\bar \mu_h^{(1)})$, at the four available $\beta$ values. We have considered the following (phenomenological) linear fit which, as shown in fig.~\ref{fig:Mmul}, turns out to describe well the lattice data,
\be
M_{hl}(\bar \mu_h^{(1)}) = C_1 + C_2 \, \bar \mu_l + C_3 \, a^2 \ .
\ee
%%%%%%%%%%%%%%%%%%%%%%%%%%%%%%%%%%%%%%%%%%%%%%%%%%%%%
%%%%%%%%%%%%%%%%%%%%%%%%%%%%%%%%%%%%%%%%%%%%%%%%%%%%%

\begin{figure}[t]
\begin{center}
\includegraphics[width=0.52\textwidth]{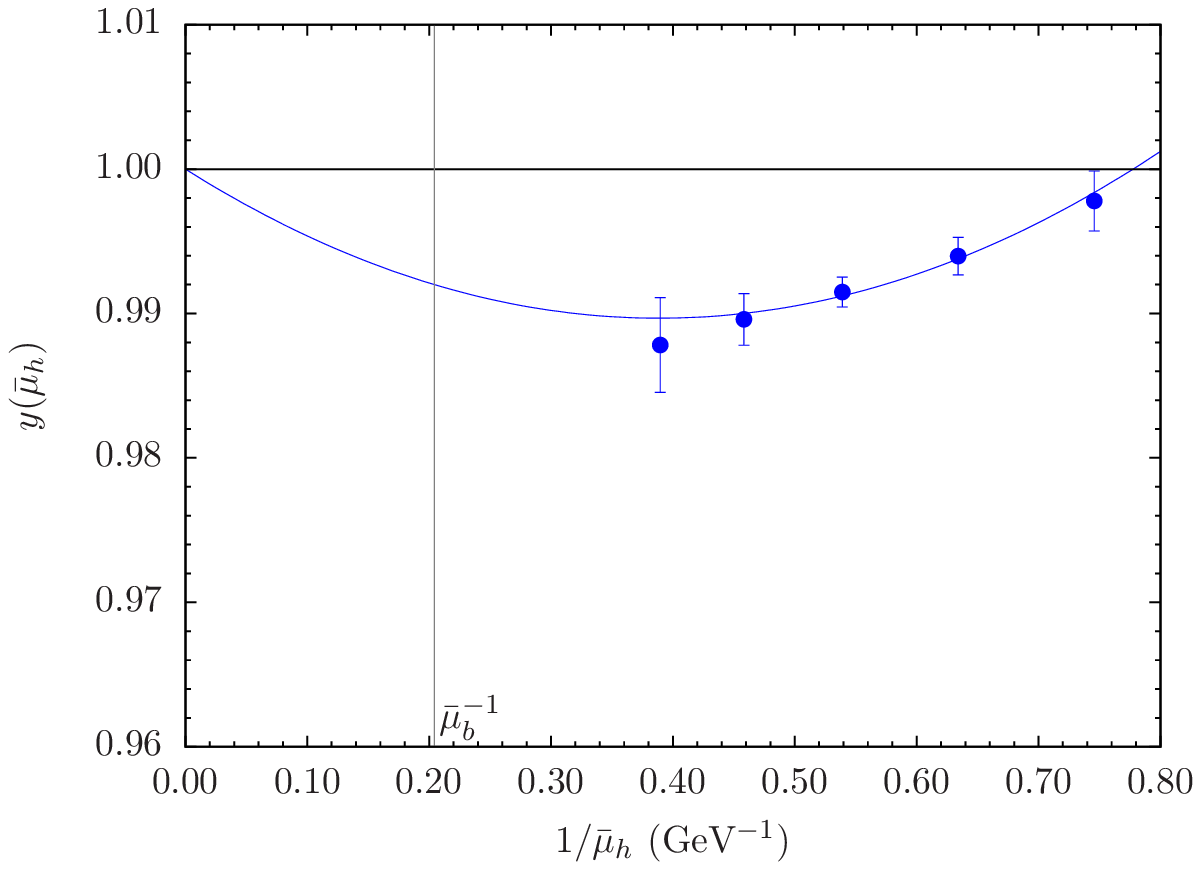}
\caption{\sl 
Heavy quark mass dependence of the ratio $y(\bar\mu_h)$ extrapolated to the physical value of the light quark mass and to the continuum limit. The vertical line represents the value of the physical $b$ quark mass.}
\label{fig:Mratiomuh}
\end{center}
\end{figure}

After performing the continuum and chiral extrapolation of the ratios~(\ref{RATM}), we study their dependence on the inverse heavy quark mass. Inspired by HQET, we perform a polynomial fit in $1/\bar\mu_h$, of the form
\be
y(\bar\mu_h) = 1 + \frac{\eta_1}{\bar\mu_h} +  \frac{\eta_2}{\bar\mu_h^2} \, ,
\label{muhfit}
\ee
which imposes the constraint $y=1$ at the static point. The fit is illustrated in fig.~\ref{fig:Mratiomuh}.
A detailed discussion of the $\mu_h$ dependence of the ratio $y$, in comparison to the HQET expectation, is provided in the Appendix.
%%%%%%%%%%%%%%%%%%%%%%%%%%%%%%%%%%%%%%%%%%%%%%%%%%%%%
%%%%%%%%%%%%%%%%%%%%%%%%%%%%%%%%%%%%%%%%%%%%%%%%%%%%%

The value of the $b$ quark mass is finally determined by considering the following equation
\be
y(\bar\mu_h^{(2)})\, y(\bar\mu_h^{(3)})\,\ldots \, y(\bar\mu_h^{(K+1)})=\lambda^{-K} \,
\frac{M_{hu/d}(\bar\mu_h^{(K+1)})}{M_{hu/d}(\bar\mu_h^{(1)})} \cdot
\Big{[}\frac{\rho( \bar\mu_h^{(1)},\mu^*)}{\rho( \bar\mu_h^{(K+1)},\mu^*)}\Big{]}\, , 
\label{ITER}
\ee
which should be looked at as a relation between the mass of the heavy-light meson, $M_{hu/d}(\bar\mu_h^{(K+1)})$, and the corresponding heavy quark mass $\bar\mu_h^{(K+1)}$, being $M_{hu/d}(\bar\mu_h^{(1)})$ the initial triggering value. The $b$ quark mass is then determined by finding the value of $K$ at which $M_{hu/d}(\bar\mu_h^{(K+1)})$ takes the experimental value of the $B$-meson mass, $M_B$. Calling $K_b$ the solution of the resulting eq.~(\ref{ITER}) (we find $K_b=9$), 
one gets for $\bar\mu_b=\mbar_b(2 \gev)$ the simple relation
\be
\bar\mu_b=\lambda^{K_b}\bar\mu_h^{(1)}=4.91(15)\,\gev\, .
\label{MB}
\ee
We observe that it is always possible to guarantee that the solution $K_b$ is an integer number through a slight variation of the parameter $\lambda$ and/or of the triggering mass $\bar\mu_h^{(1)}$. 

An equivalent method consists in determining the $b$ quark mass by studying $M_{hs}$ instead of $M_{hu/d}$ and using in input the experimental $B_s$-meson mass value, $M_{Bs}=5.366\,\gev$. A very similar result is obtained from this analysis:  $\bar\mu_b=4.92(13)\,\gev$. The small difference ($0.01\,\gev$) with respect to eq.~(\ref{MB}) indicates a good control of the chiral extrapolation which, in particular, in the heavy-strange meson case involves only the sea quark mass. The main effect of the uncertainty due the chiral extrapolation is accounted for by the error quoted in eq.~(\ref{MB}), which comes from the chiral, continuum and $1/\mu_h$ fits.

In order to estimate the residual uncertainty due to discretization effects, we have tried to include in the continuum extrapolation, besides the leading ${\cal O}(a^2)$ correction, an additional $a^4$ term. We find, however, that this subleading contribution cannot be fitted with our data. Therefore, we have repeated the analysis by excluding the data at the coarsest lattice spacing ($\beta=3.80$). The difference in the determination of the $b$ quark mass turns out to be of $0.05\,\gev$.

In order to estimate the systematic error associated to the interpolation of $y(\bar \mu_h)$ as a function of $1/\bar\mu_h$, we have repeated the whole analysis by choosing a third order polynomial in $1/\bar\mu_h$ (rather than a second order
one, as in the ansatz~(\ref{muhfit})). This change resulted in an increase of the b quark
mass of about $0.5\%$, corresponding to a shift of $\simeq 0.02$ GeV of the
central value result of eq.~(\ref{MB}).  

An additional uncertainty is introduced by the truncation of the perturbative series in the determination of the pole mass in eq.~(\ref{PMA}), which is affected by renormalon ambiguities. When comparing the results obtained with the NLO definition of the pole mass to the results found with the LO one, the difference in the $b$ quark mass is found to be small, of about $0.01\,\gev$. The sensitivity to the pole mass definition, which appears in the intermediate steps, thus largely cancels out in the final determination.

We finally quote the $b$ quark mass at the conventional renormalization scale of $m_b$ itself
\be
\bar m_b (m_b)=4.29(13)(4)\,\gev\,=4.29(14)\,\gev\, ,
\label{MBmb}
\ee
where the first error is of statistical and fitting origin and the second one is the sum in quadrature of the residual systematic uncertainties discussed above. The present result for the $b$ quark mass has a central value which is smaller by approximately one standard deviation than the value found in~\cite{Blossier:2009hg}, and an uncertainty which is reduced by almost a factor two, reflecting the various improvements implemented in the present analysis.

\subsection{Decay constants}
A similar strategy is employed by applying the ratio method to determine the $B$ and $B_s$ meson decay constants. The HQET asymptotic prediction for the decay constant is
\be
\lim_{\mu_h^{\rm{pole}}\to \infty} f_{h\ell} \sqrt{\mu_h^{\rm{pole}}}=\mbox{constant} \, .
\label{FB}
\ee
Therefore, the ratios with static limit equal to one of interest in this case are, for $f_B$ and $f_{Bs}$,~\cite{Blossier:2009hg}
\bea
z(\bar\mu_h,\lambda;\bar\mu_\ell, a)&\equiv&  
\lambda^{1/2} \frac{f_{h\ell}(\bar\mu_h,\bar\mu_\ell, a)}{f_{h\ell}(\bar\mu_h/\lambda,\bar\mu_\ell, a)}
\cdot \frac{C^{stat}_A(\mu_b^*,\bar\mu_h/\lambda)}{C^{stat}_A(\mu_b^*, \bar\mu_h)} 
\frac{[\rho( \bar\mu_h,\mu^*)]^{1/2}}{[\rho( \bar\mu_h/\lambda,\mu^*)]^{1/2}} \nn \\
z_s(\bar\mu_h,\lambda;\bar\mu_\ell,\bar\mu_s, a)&\equiv& 
\lambda^{1/2} \frac{f_{h s}(\bar\mu_h,\bar\mu_\ell,\bar\mu_s, a)}{f_{h s}(\bar\mu_h/\lambda,\bar\mu_\ell,\bar\mu_s, a)}
\cdot \frac{C^{stat}_A(\mu_b^*,\bar\mu_h/\lambda)}{C^{stat}_A(\mu_b^*,\bar\mu_h)} 
\frac{[\rho( \bar\mu_h,\mu^*)]^{1/2}}{[\rho( \bar\mu_h/\lambda,\mu^*)]^{1/2}} \,.
\label{RATMZ}
\eea
The ratio of $\rho$ factors (raised to the appropriate power) is present to convert $\overline{MS}$ heavy quark masses to pole masses as in eq.~(\ref{RATM}).
The factor $C^{stat}_A(\mu_b^*,\bar\mu_h)$, defined as
\be
\Phi_{h s}(\mu_b^*)=\left[ C_A^{stat}(\mu_b^*,\bar\mu_h)\right]^{-1}\,\cdot \Phi_{h s}^{\rm QCD}(\bar\mu_h)\,,
\label{eq:match}
\ee
provides the matching between the decay constant in QCD for a heavy quark mass $\bar \mu_h$ and in HQET, and the running of the static axial current to the renormalization scale $\mu_b^*$, and it is known up to N$^2$LO  in PT~\cite{ZAPT}.
The NLO expression used in the present analysis reads
\bea
C^{stat}_A(\mu_b^*,\bar\mu_h)&=&\Biggl(\frac{\alpha^{\overline{\mbox{MS}}}(\bar \mu_h)}{\alpha^{\overline{\mbox{MS}}}( \mu_b^*)}\Biggr)^{-\frac{6}{33-2\,N_f}}\,\cdot
\Biggl[1-\Biggl(\frac{-3951+300\,N_f+60\,N_f^2+(924-56\,N_f)\pi^2}{9(33-2\,N_f)^2} \Biggr)\,\Biggr.\nn\\
&&\Biggl. \cdot\frac{\alpha^{\overline{\mbox{MS}}}(\bar \mu_h)-\alpha^{\overline{\mbox{MS}}}(\mu_b^*)}{4\pi}\Biggr]\cdot
\Biggl[1-\frac{8}{3}\frac{\alpha^{\overline{\mbox{MS}}}(\bar \mu_h)}{4\pi}\Biggr]\,,
\label{eq:CA}
\eea
with $N_f=2$.

In order to have better control on the chiral extrapolation, we consider as primary quantities in the present analysis the decay constant $f_{B_s}$, whose dependence on the light quark mass only occurs through sea effects, and the ratio $f_{B_s}/f_B$ which provides a direct determination of the SU(3) breaking effect in the decay constant. Within the ratio method, these quantities are obtained from the ratio $z_s$ and the double ratio $z_s/z$.

%%%%%%%%%%%%%%%%%%%%%%%%%%%%%%%%%%%%%%%%%%%%%%%%%%%%%
\begin{figure}[t]
\includegraphics[width=0.48\textwidth]{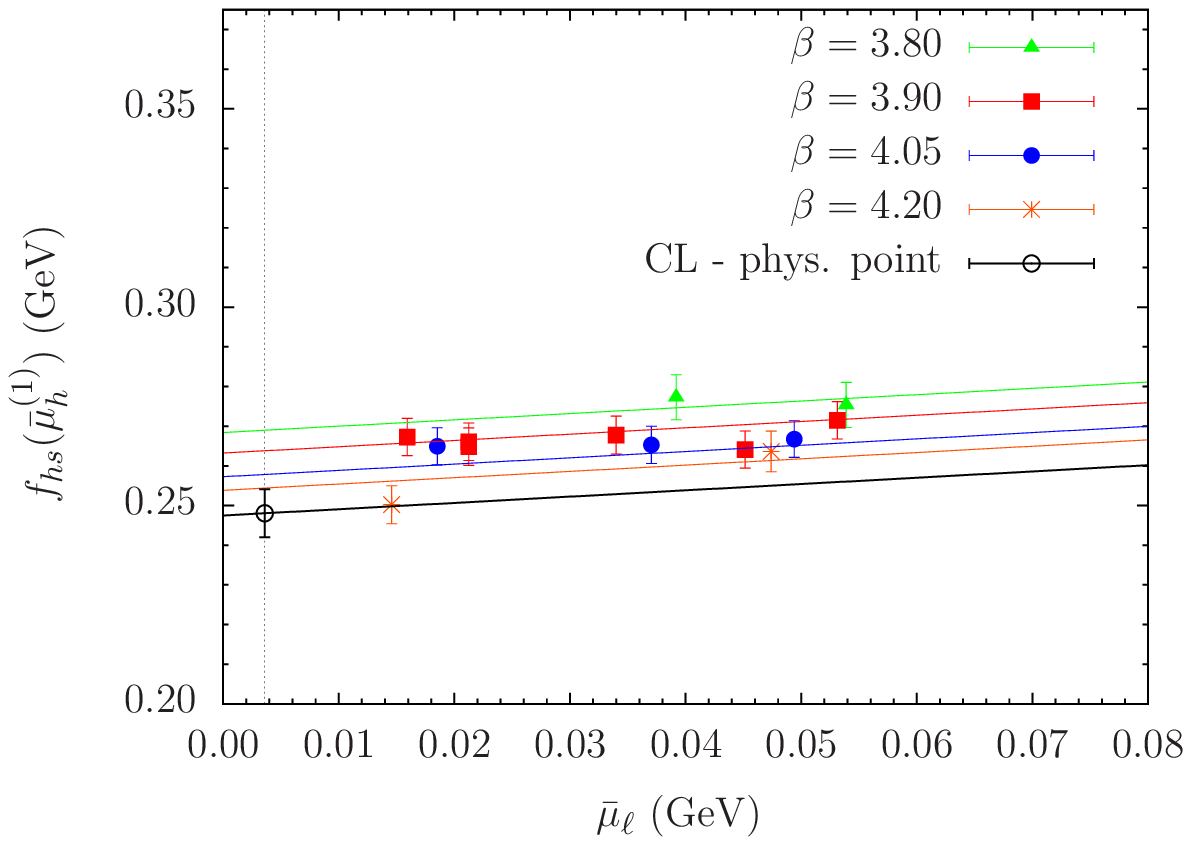}
\includegraphics[width=0.48\textwidth]{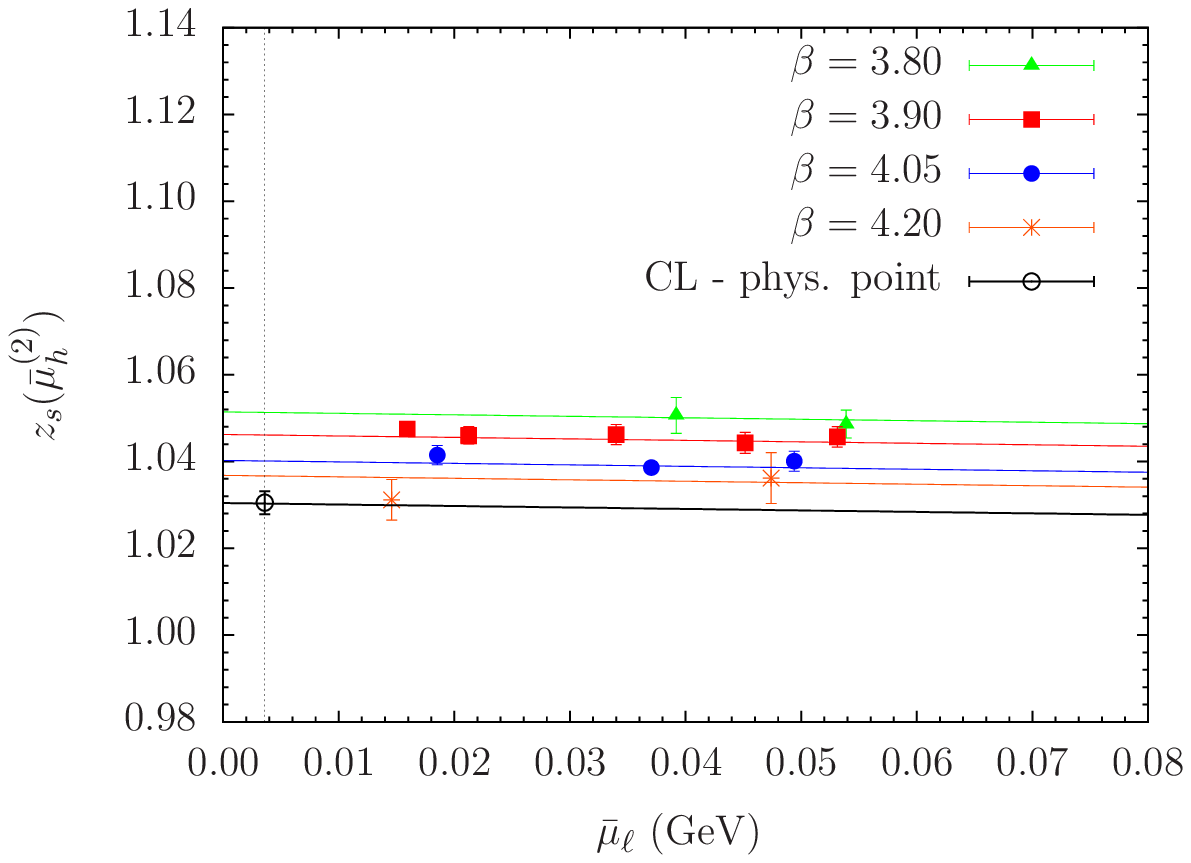}
\caption{\sl Light quark mass dependence of the decay constant $f_{hs}(\bar\mu_h^{(1)})$ (left) and of the ratio $z_s(\bar\mu_h^{(2)})$ (right) at the four values of the lattice spacing.}
\label{fig:ratiofmul}
\vspace{0.4cm}
\end{figure}
%%%%%%%%%%%%%%%%%%%%%%%%%%%%%%%%%%%%%%%%%%%%%%%%%%%%%
\begin{figure}[t!]	
\includegraphics[width=0.48\textwidth]{fhsfhl_ml.eps}
\includegraphics[width=0.48\textwidth]{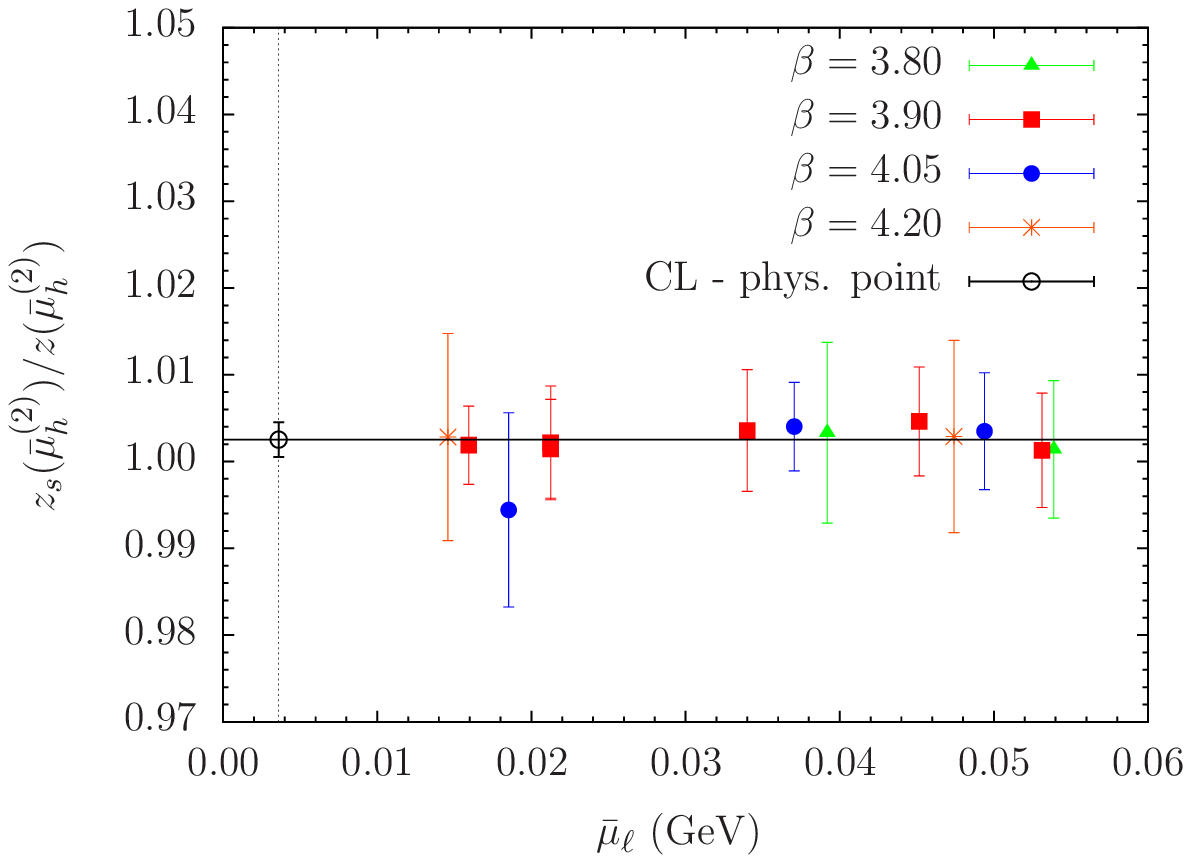}
\caption{\sl Light quark mass dependence of the ratio of decay constants $f_{hs}(\bar\mu_h^{(1)})/f_{h l}(\bar\mu_h^{(1)})$ (left) and of the double ratio $z_s(\bar\mu_h^{(2)})/z(\bar\mu_h^{(2)})$ (right) at the four values of the lattice spacing.}
\label{fig:doubleratiofmul}
\end{figure}
%%%%%%%%%%%%%%%%%%%%%%%%%%%%%%%%%%%%%%%%%%%%%%%%%%%%%
Both $z_s$ and $z_s/z$ have a smooth chiral and continuum limit, as illustrated in figs.~\ref{fig:ratiofmul} and~\ref{fig:doubleratiofmul}. In particular, the results for the double ratio $z_s/z$ turn out to be well described by both a linear and a constant behavior in both $\bar\mu_\ell$ and $a^2$ (see fig.~\ref{fig:doubleratiofmul} right).
For simplicity reasons the constant fit ansatz was chosen.
In the left panels of figs.~\ref{fig:ratiofmul} and~\ref{fig:doubleratiofmul} we show the chiral and continuum extrapolation of $f_{hs}$ and $f_{h s}/f_{h \ell}$ at the initial (triggering) mass $\bar\mu_h^{(1)}$. For $f_{h s}/f_{h \ell}$, heavy meson chiral perturbation theory (HMChPT) predicts at the NLO a linear+logarithmic dependence on the light quark mass, since a chiral log controls the chiral behavior of $f_B$ (see eq.~(\ref{eq:HMChPTB}) below). With our results, the logarithmic dependence cannot be appreciated, and we thus perform also a simpler linear fit in the light quark mass which turns out to describe well the lattice data. As discussed in section 5, we eventually average the results obtained from the HMChPT and the linear fits and include the difference in the systematic uncertainty. For $f_{h s}$, which depends on the light quark mass for sea effects only, we have implemented both a linear and a quadratic fit. They turn out to provide essentially identical results.

%%%%%%%%%%%%%%%%%%%%%%%%%%%%%%%%%%%%%%%%%%%%%%%%%%%%%
 \begin{figure}[t]	
  \includegraphics[width=0.48\textwidth]{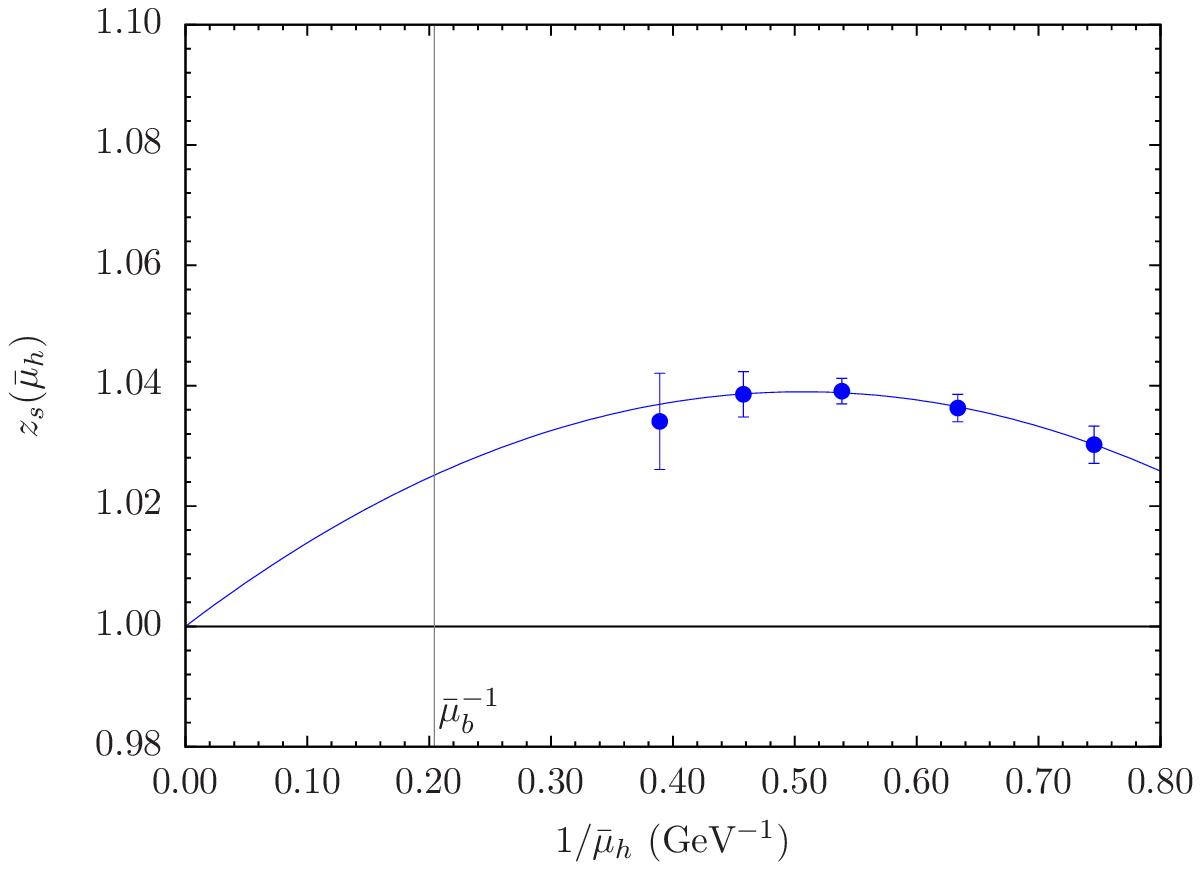}
  \includegraphics[width=0.48\textwidth]{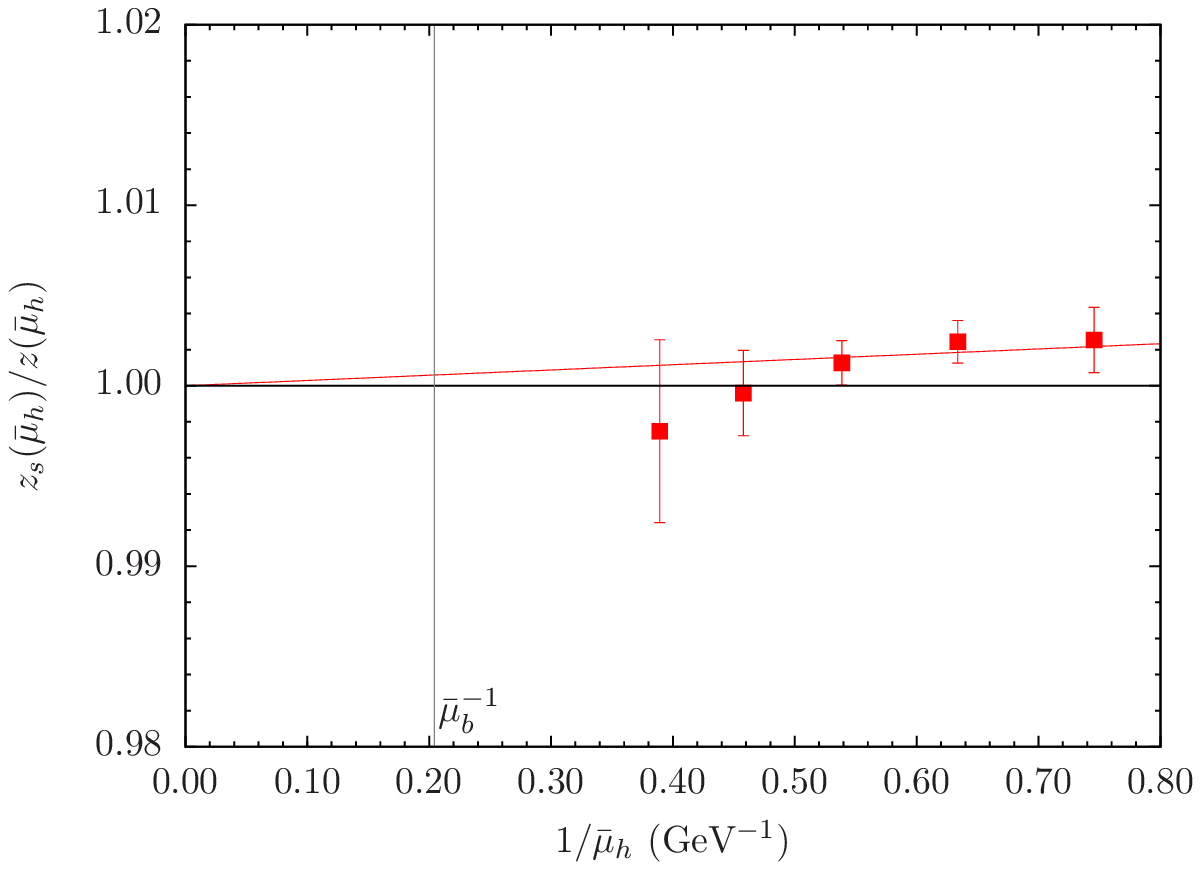}
   \caption{\sl Heavy quark mass dependence of the ratio $z_s(\bar\mu_h)$ (left) and of the double ratio $z_s(\bar\mu_h)/z(\bar\mu_h)$ (right) extrapolated to the physical value of the light and strange quark masses and to the continuum limit.  The vertical line represents the value of the physical $b$ quark mass.}
\label{fig:fratiomuh}
\end{figure}
%%%%%%%%%%%%%%%%%%%%%%%%%%%%%%%%%%%%%%%%%%%%%%%%%%%%%
Finally, we study the dependence of the ratio $z_s$ and the double ratio $z_s/z$ on the heavy quark mass, which are shown in fig.~\ref{fig:fratiomuh}. For $z_s$ we perform a quadratic interpolation to the $b$ quark mass as for the ratio $y(\bar\mu_h)$ in eq.~(\ref{muhfit}) and, also in this case, a detailed discussion is provided in the Appendix. For $z_s/z$, the dependence on the heavy quark mass is barely visible, so that in this case we perform either a linear interpolation in $1/\bar\mu_h$ or we fix this ratio equal to its asymptotic heavy-quark mass limit, $z_s/z=1$.

\section{Interpolation method}
As already mentioned, the interpolation method consists in interpolating to the $b$ quark mass the relativistic results obtained for values of the heavy quark masses in the range around and above the physical charm (up to twice to three times its value) and the result evaluated in the static limit by simulating the HQET on the lattice. In this section, we describe these results by addressing, in turn, the calculation with relativistic lattice QCD in the charm mass region, the calculation within the HQET on the lattice, and the interpolation among the two sets of results.

\subsection{Decay constants in relativistic QCD}
The lattice relativistic data for the heavy-light and heavy-strange meson masses and decay constants are the same used for the ratio method. We considered in the analysis four values of the lattice spacing and the values of valence quark masses collected in Table~\ref{tab:val}. With respect to the preliminary results with this method presented in~\cite{Blossier:2009gd}, we added an ensemble with a lighter quark mass at $\beta=4.2$ and, for other ensembles, we increased the statistics. Another update w.r.t to the analysis in~\cite{Blossier:2009gd} concerns the renormalization constants, which had preliminary values at the time of~\cite{Blossier:2009gd}, and have been later updated and published in~\cite{Constantinou:2010gr}.
The main improvement, however, concerns the disentanglement of the heavy mass dependence from discretization effects. In the present analysis the extrapolation to the continuum limit is performed at fixed (renormalized) heavy quark mass. The whole analysis consists in the following steps.

First, we slightly interpolate the lattice data to reach a set of reference heavy quark masses equal at the four $\beta$ values. This allows us to study discretization effects at fixed heavy quark mass. We have chosen the reference masses within the range of the simulated values. In the $\msb$ scheme, at $\mu=2 \gev$, the set of reference masses is $\bar \mu_h = \{1.25, 1.50, 1.75, 2.00, 2.25\}\gev$.

We have then performed a combined continuum and chiral extrapolation, at fixed reference heavy quark mass.
As for the ratio method, we consider as primary quantities $f_{h s}$ and the ratio $f_{h s}/f_{h \ell}$ obtained in the present analysis from $\Phi_{h s}$ and $\Phi_{h s}/\Phi_{h \ell}$, where
\be
\Phi_{h s} = f_{h s} \sqrt{M_{h s}} \qquad {\rm and} \qquad 
\frac{\Phi_{h s}}{\Phi_{h \ell}} = \frac{f_{h s}}{f_{h \ell}} \sqrt{\frac{M_{h s}}{M_{h \ell}}} \, .
\ee
An alternative analysis based on the definition of $\Phi_{h \ell(s)}$ in terms of the pole mass, rather than the meson mass, i.e. $\Phi_{h \ell(s)} = f_{h \ell(s)} \sqrt{\mu_b^{\rm pole}}$, has been also performed, leading to fully equivalent results. 

The light quark mass dependence predicted for $\Phi_{h \ell}$ and $\Phi_{h s}$ by HMChPT~\cite{GoityTP, GrinsteinQT, Sharpe:1995qp} at the NLO reads
\bea
&\Phi_{h \ell}(a,\bar \mu_\ell, \bar \mu_h)=A_h\,\left[1-\dfrac{3(1+3 \hat g^2)}{4} \dfrac{2\,B_0\,\bar \mu_\ell}{(4\,\pi\,f_0)^2}\log\left(\dfrac{2\,B_0\,\bar \mu_\ell}{(4\,\pi\,f_0)^2}\right)+B_h\,\bar \mu_\ell+C_h\,a^2\right]\,,& \nn \\
&&\nn\\
&\Phi_{h s}(a,\bar \mu_\ell, \bar \mu_s, \bar \mu_h)=D_h\,\left(1+E_h\,\bar \mu_\ell+F_h\,\bar \mu_s+G_h\,a^2\right)\,,&
\label{eq:HMChPTB}
\eea
where we have also included in the above expressions a linear dependence on $a^2$ to account for leading discretization effects. The subscript $h$ in the fit parameters of eq.~(\ref{eq:HMChPTB}) denotes the dependence on the heavy quark mass.

As previously discussed for the ratio method, the contribution of chiral logs in the ratio $\Phi_{h s}/\Phi_{h \ell}$, predicted by HMChPT, cannot be appreciated with our data (see fig.~\ref{fig:doubleratiofmul} left). Thus, in order to perform the chiral extrapolation to the physical light quark mass, we also perform a simple linear fit in $\bar \mu_\ell$ and eventually take the average of the two results. In the fit based on HMChPT, we take for the parameter $\hat g$ the value  $\hat g=0.61(7)$ obtained from the experimental measurement of the $g_{D^* D \pi}$ coupling~\cite{Nakamura:2010zzi}.
We choose this value, instead of the HQET prediction $\hat g=0.44(8)$~\cite{OhkiPY, Becirevic:2009yb}, as we fit data that are close to the charm mass region and in order to conservatively include in the average the maximum spread resulting from the different ways of performing the chiral extrapolation of our data. For $\Phi_{h s}$, as for $f_{hs}$ within the ratio method (see fig.~\ref{fig:ratiofmul} left), we have tried both a linear and a quadratic fit in $\bar \mu_\ell$, obtaining very similar results.

%%%%%%%%%%%%%%%%%%%%%%%%%%%%%%%%%%%%%%%%%%%%%%%%%%%%%%%%%
\begin{figure}[t]
\begin{center}
\includegraphics[width=0.5\textwidth]{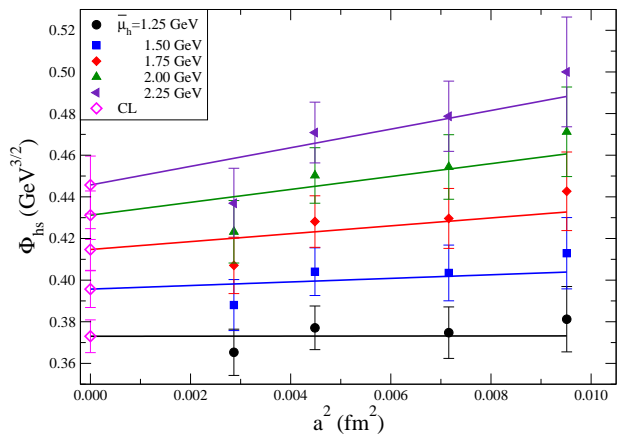}
\end{center}
\caption{\sl Dependence of $\Phi_{h s}$ on the squared lattice spacing, at the five  reference values of the heavy quark mass (for a fixed value of the up/down and strange quark masses).}
\label{fig:Phia2}
\end{figure}
%%%%%%%%%%%%%%%%%%%%%%%%%%%%%%%%%%%%%%%%%%%%%%%%%%%%%%%%%
For illustration, the size of discretization effects in the calculation of $\Phi_{h s}$ is shown in fig.~\ref{fig:Phia2}, for a simulated value of the light quark mass of about $50\,\mev$ and with $\bar \mu_s \approx \bar \mu_s^{phys}$. It is interesting to note that lattice artifacts turn out to be small. We find that this is a consequence of a partial cancellation between discretization terms in the decay constant and in the rooted meson mass, which are of similar size and opposite sign. For the same reason, the ratio $\Phi_{h s}/\Phi_{h \ell}$ turns out to be practically independent of the lattice spacing.

\subsection{Decay constants in the static limit of HQET}
We now summarize the procedure adopted to evaluate the static-light meson masses and decay constants, which follows and improves the analysis of~\cite{Blossier:2009gd}. We use the same setup as for the recent ETMC study of the static-light meson spectrum \cite{Jansen:2008si,:2010iv}. Technical details regarding e.g.\ number of ETMC $N_f=2$ gauge configurations considered, meson creation operators, corresponding correlation matrices and their analysis, smearing techniques to enhance the signal quality~\cite{DellaMorte:2005mn} as well as efficient propagator computation can be found in these references.

The lattice action used to describe the static quark is the HYP2 static action~\cite{DellaMorte:2005mn,Hasenfratz:2001hp,DellaMorte:2003mn}
\be
S_h= a^4\,\sum_x \bar \psi_h (x) D_0\, \psi_h (x)
 = a^3\,\sum_x \bar \psi_h (x)  [\psi_h(x)-V_{\rm HYP}^\dagger (x-a \hat 0,0)\, \psi_h(x-a \hat 0)] \,,
\ee
where $V_{\rm HYP}$ is the so-called HYP-link, which is a gauge covariant function of the gauge links located within a hypercube.

Static-light correlators have been calculated for a subset of the configuration ensembles used for the relativistic calculation, namely for two values of the lattice spacing, $\beta=3.90$ and $\beta=4.05$.

Due to the mixing between the pseudoscalar and the scalar currents occurring in the static-light framework, $\Phi^{\rm stat}_{B (s)}$ is obtained as a linear combination of two matrix elements, 
\be
\Phi^{\rm stat}_{B_q} = Z_P^{\rm stat} \langle 0 | \bar{\psi}_{\rm h}\gamma_5\chi_{q} | B^{\rm stat}_{(q)} \rangle+ i\,r_q\,Z_S^{\rm stat} \langle 0 | \bar{\psi}_{\rm h}\chi_{q} | B^{\rm stat}_{(q)} \rangle \,,
\ee
where $\chi_{q}=e^{-i r_q (\pi/4) \gamma_5} \psi_{q}$ is the light quark field in the twisted basis, $r_q=\pm1$ is the corresponding Wilson parameter, and $Z_P^{\rm stat}$ and $Z_S^{\rm stat}$ the static-light renormalization constants. In order to improve the signal quality of the lattice data we have combined the results with $r_q=+1$ and $r_q=-1$, which are identical on average due to twisted mass symmetries.

In order to check the stability of the fitting method to extract the above matrix elements, we have performed, as in~\cite{Jansen:2008si}, the computations with different parameters: number of states to be fitted in correlators, fitting time ranges, operator content of the correlation matrices. We have obtained results that are consistent within statistical errors.

For our setup (tree-level improved Symanzik gauge action and HYP2 static action) the non-perturbative values of the static-light renormalization constants are not available at present. We thus rely on the perturbative estimate at one loop~\cite{Blossier:2011dg} evaluated using a boosted coupling. It provides $Z_P^{\rm stat}=\{0.85(8), 0.86(7)\}$ and $Z_S^{\rm stat}=\{0.93(4), 0.94(3)\}$, at $\beta=\{3.90, 4.05\}$ respectively, in the $\msb$ scheme at the renormalization scale $1/a$. The uncertainty on $Z^{\rm stat}_{P,S}$ has been conservatively estimated as half of the deviation from unity.

%%%%%%%%%%%%%%%%%%%%%%%%%%%%%%%%%%%%%%%%%%%%%%%%%%%%%%%%%
\begin{figure}[t]	
\includegraphics[width=0.42\textwidth,angle=270]{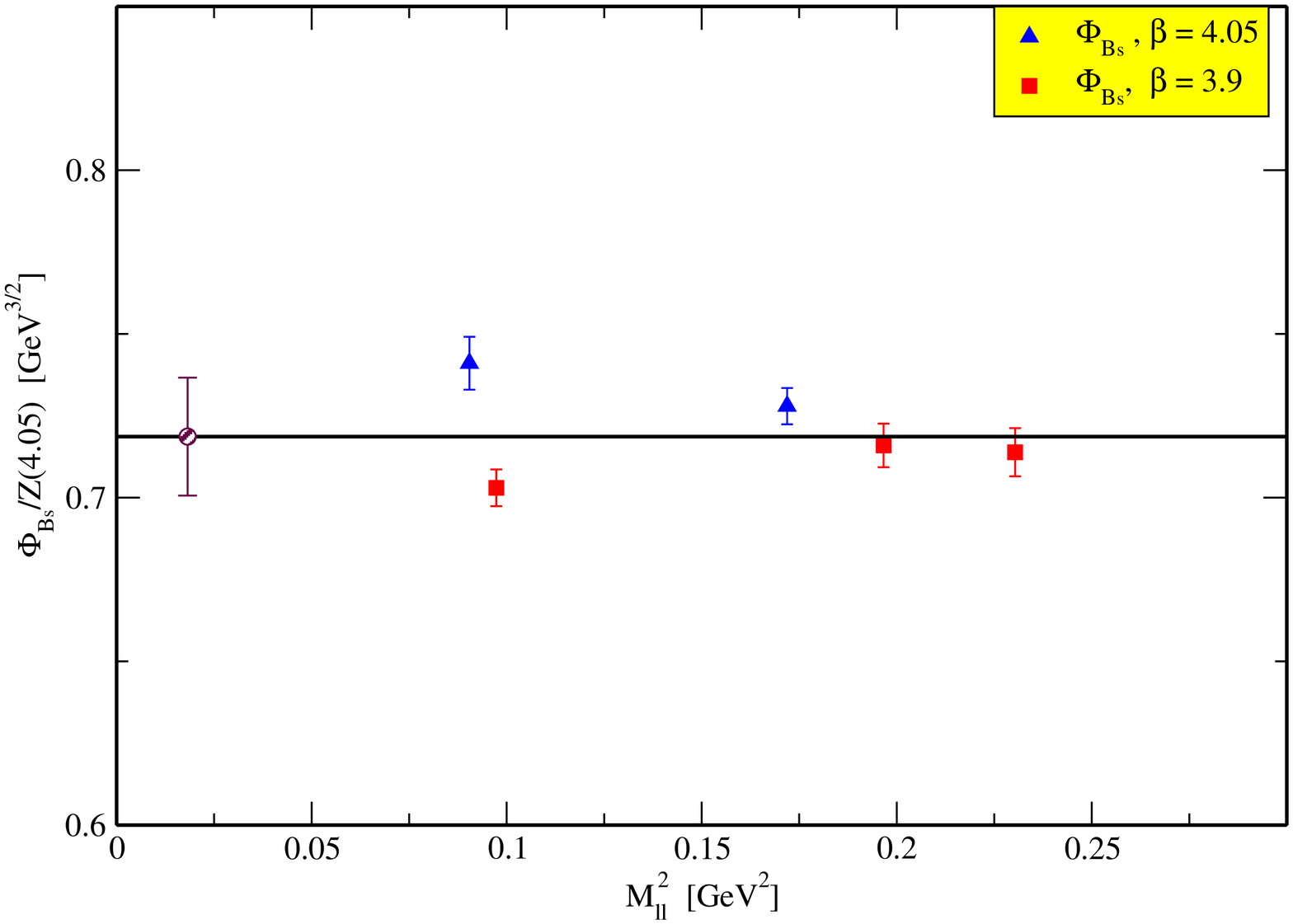}
\includegraphics[width=0.42\textwidth,angle=270]{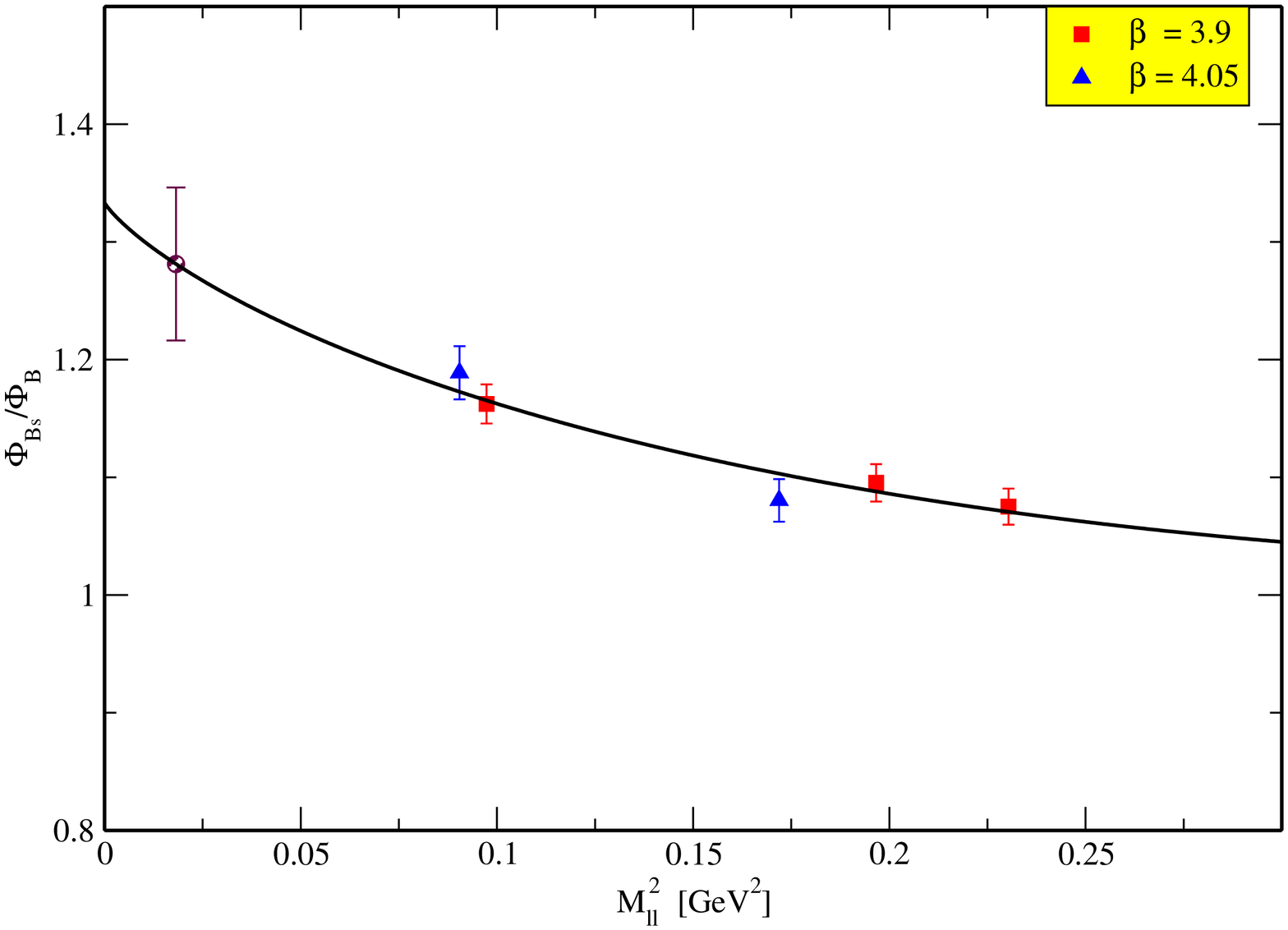}
\caption{\sl The combination $\Phi^{\rm stat}_{Bs}/Z (4.05)$ (left), with $Z \equiv \left(Z_P^{\rm stat} + Z_S^{\rm stat}\right)/2$, and the ratio $\frac{\Phi^{\rm stat}_{B_s}}{\Phi^{\rm stat}_B}$ (right) as a function of the (squared) light pseudoscalar meson mass. In the left plot the data at $\beta = 3.90$ have been multiplied by the appropriate factor to match the same scale for the data at $\beta=4.05$.}
\label{fig:chiralfit}
\end{figure}
%%%%%%%%%%%%%%%%%%%%%%%%%%%%%%%%%%%%%%%%%%%%%%%%%%%%%%%%%
The chiral extrapolation of $\Phi^{\rm stat}_{Bs}$ and $\Phi^{\rm stat}_{Bs}/\Phi^{\rm stat}_B$ has been performed, as for the relativistic data, using the HMChPT functional forms given in eq.~(\ref{eq:HMChPTB}), and it is shown in fig.~\ref{fig:chiralfit} in terms of the light meson mass squared $M_{ll}^2$. In this case, however, discretization terms are set to zero, as the fit includes data at only two values of the lattice spacing.

The matrix element shown in the left plot of fig.~\ref{fig:chiralfit} is chosen to be, for better clarity, the ratio $\Phi^{\rm stat}_{Bs}/Z$, where $Z = (Z_S^\textrm{stat} + Z_P^\textrm{stat}) / 2$. This quantity only depends on the ratio of renormalization constants, $Z_S^\textrm{stat}/Z_P^\textrm{stat}$, but not on their average. One can also show that the relative systematic errors due to inaccurate knowledge of $Z_S^\textrm{stat} / Z_P^\textrm{stat}$ on these matrix elements are $\Delta(Z_S^\textrm{stat} / Z_P^\textrm{stat}) \times \mathcal{O}(a)$, where $\Delta(Z_S^\textrm{stat} / Z_P^\textrm{stat})$ is the error on $Z_S^\textrm{stat} / Z_P^\textrm{stat}$ and $\mathcal{O}(a)$ denotes matrix elements which are proportional to the lattice spacing. % and have numerically been found to be around $0.1 \ldots 0.2$ for our ensembles. 
Consequently, the matrix elements shown in fig.~\ref{fig:chiralfit} are only slightly affected by the systematic uncertainty on the renormalization constants due to the use of one-loop lattice perturbation theory.
Of course the evaluation of $\Phi^{\rm stat}_{Bs}$ at each lattice spacing requires the info on one renormalization constant, which is conveniently chosen as $(Z_S^\textrm{stat} + Z_P^\textrm{stat}) / 2$.

Fig.~\ref{fig:chiralfit} shows that the chiral dependence on the light pseudoscalar mass is found to be very smooth for $\Phi^{\rm stat}_{Bs}$, and a simple constant fit is used to perform the chiral extrapolation. In order to estimate the uncertainty due to the chiral extrapolation, a linear fit is tried for comparison and the difference is included in the systematic error. 
For the ratio $\Phi^{\rm stat}_{Bs}/\Phi^{\rm stat}_B$, besides the fit based on HMChPT we have also considered both a linear (in $M^2_{ll}$) and a quadratic fit. The variation of the result is found to be of the order of 5\% and it is included in the systematic uncertainty.

It is clear from fig.~\ref{fig:chiralfit} (right) that the data for the ratio $\Phi^{\rm stat}_{Bs}/\Phi^{\rm stat}_B$ are affected by negligible cut-off effects. Some discretization effects are visible, instead, in the left plot of fig.~\ref{fig:chiralfit} which shows the results for $\Phi^{\rm stat}_{Bs}$. In order to estimate their magnitude we compare the results for $\Phi^{\rm stat}_{Bs}$ obtained by either fitting together data at $\beta = 3.90$ and $\beta = 4.05$ or fitting data at one lattice spacing only. The difference, which turns out to be at the level of 3\%, is included in the systematic uncertainty.

For $\Phi^{\rm stat}_{Bs}$ the simulated (bare) strange quark mass is fixed to the values $a\mu_s=0.0220$ at $\beta=3.9$ and $a\mu_s=0.0170$ at $\beta=4.05$ respectively. These values correspond to the physical strange mass, as obtained from the analysis at fixed lattice spacing. The continuum limit performed in the study of~\cite{Blossier:2010cr} has later provided a value for the strange quark mass which is smaller by approximately 22\% at $\beta=3.9$ and 13\% at $\beta=4.05$. The effect of this mismatch is discussed in the following section and included in the systematic uncertainty. 

As a last step in the analysis of the static data, a perturbative evolution at NLO~\cite{ZAPT} has been applied to evolve the results in the $\overline{MS}$ scheme from the initial scale $\mu=1/a$ to a common reference scale $\mu_b^* = 4.5 \gev$, obtaining
\be
\Phi^{\rm stat}_{Bs}(\mu_b^*)=0.67(4)\gev^{3/2} \qquad \mathrm{and} 
\qquad \Phi^{\rm stat}_{Bs}/\Phi^{\rm stat}_{B}=1.28(7)\, .
\label{phistat}
\ee
For the $B$ meson, eq.~(\ref{phistat}) corresponds to the result
\be
\Phi^{\rm stat}_{B}(\mu_b^*)=0.52(3)\gev^{3/2}\,.
\ee

As a further consistency check between the analyses based on the ratio and the interpolation methods, we have used the ratio method to predict the value of $\Phi_{Bs}/\Phi_{B}$ in the static limit, finding $\Phi^{\rm stat}_{Bs}/\Phi^{\rm stat}_{B}=1.20(5)$. The latter is compatible with the result in eq.~(\ref{phistat}) obtained from the direct lattice simulation in the HQET. 

\subsection{Interpolation of relativistic and static data}
In order to perform a combined fit of relativistic and static data, we convert the relativistic values of $\Phi_{h s}$ from QCD to HQET, by using the NLO matching and evolution factor $C_A^{stat}$ (see eqs.(\ref{eq:match}) and~(\ref{eq:CA})).
The renormalization scale is chosen to be $\mu_b^*=4.5\,\gev$ as for the static data.
Note that in the ratio $\Phi_{h s}/\Phi_{h \ell}$ the $C_A^{stat}$ factor cancels out.

\begin{figure}[t]
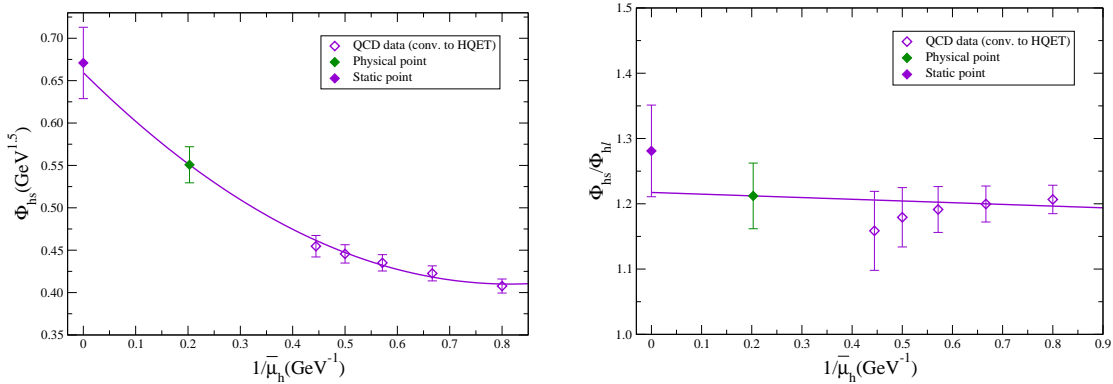
	
\includegraphics[width=0.43\textwidth]{estrapolazione_fBs.eps}
\hspace{0.5cm}
\includegraphics[width=0.43\textwidth]{estrapolazione_fBsfB.eps}
\caption{\sl Dependence of $\Phi_{h s}$ (left) and $\Phi_{h s}/\Phi_{h \ell}$ (right), in the chiral and continuum limit, on the inverse of the heavy quark mass.}
\label{fig:Phimh}
\end{figure}

The interpolation is then performed, as shown in fig.~\ref{fig:Phimh}, through a fit in $1/\bar \mu_h$, which is quadratic for $\Phi_{h s}$ (similarly to eq.~(\ref{muhfit})) and only linear for $\Phi_{h s}/\Phi_{h \ell}$, where a much smoother dependence on the heavy quark mass is found, as expected. Finally, the physical results for the decay constants are obtained by inserting the ``physical'' value of the $b$ quark mass determined from the ratio method, given in eqs.~(\ref{MB})-(\ref{MBmb}). The uncertainty on the $b$ quark mass has been propagated by assuming a gaussian distribution of the errors.
%%%%%%%%%%%%%%%%%%%%%%%%%%%%%%%%%%%%%%%%%%%%%%%%%%%%%%%%%
%%%%%%%%%%%%%%%%%%%%%%%%%%%%%%%%%%%%%%%%%%%%%%%%%%%%%%%%%

\section{Results for the decay constants}
In this section we present and discuss the final results obtained for the decay constants $f_{Bs}$, $f_{Bs}/f_B$ and $f_B$ from the ratio and the interpolation methods.

As discussed in the previous sections, in order to estimate the uncertainty due to the chiral extrapolation we compare the results of two different chiral fits. This comparison is relevant in particular for the ratio $\Phi_{h s}/\Phi_{h \ell}$, from which $\Phi_{Bs}/\Phi_{B}$ is extracted. In this case, the two fits are based either on the linear + logarithmic dependence on the light quark mass predicted by HMChPT (eq.~(\ref{eq:HMChPTB})) or on a polynomial (quadratic) behavior. For $\Phi_{h s}$ we have tried both the linear fit of eq.~(\ref{eq:HMChPTB}) and a quadratic fit. The analogous chiral fit ansatz employed in the analysis of
$f_{hs}(\bar\mu_h^{(1)})/f_{h\ell}(\bar\mu_h^{(1)})$ within the ratio
method framework have been discussed in section~3.2.
The results are collected in Table~\ref{tab:results} for both the ratio and the interpolation method. The first error quoted in the table is the one coming from the fit, and includes both the statistical error and the systematic uncertainty due to the chiral and continuum extrapolation and to the interpolation to the $b$ quark mass.
%%%%%%%%%%%%%%%%%%%%%%%%%%%%%%%%%%%%%%%%%%%%%%%%%%%%%%%%%
\begin{table}[t]
\begin{center}
\renewcommand{\arraystretch}{1.4}
\begin{tabular}{||c|c|c|c||c|c|c|c||}
\hline
\multicolumn{4}{||c||}{$f_{Bs}[\mev]$} & \multicolumn{4}{|c||}{$f_{Bs}/f_B$}  \\ \hline\hline
\multicolumn{2}{||c|}{Ratio Method} & \multicolumn{2}{|c||}{Interpol. Method} & \multicolumn{2}{|c|}{Ratio Method}  & \multicolumn{2}{||c||}{Interpol. Method} \\ \hline
Lin. & Quad. & Lin. & Quad. & HMChPT & Polyn.& HMChPT & Polyn.\\ \hline
{\small 225(7)(4)} & {\small 225(7)(4)} & {\small 237(9)(4)} & {\small 238(9)(4)}  & {\small 1.22(2)(0)} & {\small 1.14(2)(0)} & {\small 1.22(5)(2)} & {\small 1.16(6)(2)} \\ \hline
\multicolumn{2}{||c|}{225(7)(4)}& \multicolumn{2}{|c||}{238(9)(4)}& 
\multicolumn{2}{|c|}{1.18(2)(4)}& \multicolumn{2}{|c||}{1.19(5)(3)}\\ \hline
\multicolumn{4}{||c||}{232(10)}&  \multicolumn{4}{|c||}{1.19(5)}\\ \hline
\end{tabular}
\renewcommand{\arraystretch}{1.0}
\end{center}
\vspace{-0.4cm}
\caption{\sl Collection of the results obtained for $f_{Bs}$ and $f_{Bs}/f_B$ from the ratio and interpolation methods. The statistical and systematic uncertainties are summed in quadrature.
The third and fourth lines provide info on the results obtained by extrapolating to the physical pion mass point by using different chiral fit ansatz (see text). The final values, given in the last row, are an average of the results of the two methods.}
\label{tab:results}
\end{table}
%%%%%%%%%%%%%%%%%%%%%%%%%%%%%%%%%%%%%%%%%%%%%%%%%%%%%%%%%%%
The second error accounts for the additional systematic uncertainties and it has been evaluated as follows:
\begin{itemize}
\item[-] \underline{Continuum limit}: when performing the continuum limit, both in the ratio and the interpolation method, we consider a linear fit in $a^2$. Since an additional $a^4$ term cannot be fitted with our data, we estimate the uncertainty due to discretization effects by excluding data at the coarsest lattice ($\beta=3.80$). The central values for $f_{Bs}$ change by 2 and 1 MeV for the ratio and the interpolation method respectively. The corresponding changes in the values of the ratio $f_{Bs}/f_{B}$ are instead negligible.
\item[-] \underline{Heavy mass dependence}:
Within the interpolation method we estimate the uncertainty in reaching the 
physical bottom mass by including, for each $\beta$, data at two larger values of $\mu_h$, and
by choosing slightly different values for the reference masses. We find that with these
variations the central values obtained for $f_{Bs}$ change by approximately 3 MeV, while
the results for the ratio $f_{Bs}/f_B$ are practically unaffected.
In the context of the ratio method analysis in order to estimate the systematic 
error associated to the determination of $z_s(\bar\mu_b)$ we have varied the fit
ansatz by considering either a second order or a third order polynomial in $1/\bar\mu_h$.
This change produces only a 1 MeV decrease in the final value of $f_{Bs} \simeq 225$~MeV
(see Table~\ref{tab:results}). Even smaller is the relative uncertainty in the $1/\bar\mu_h$ interpolation
of the double ratio $z_s(\bar\mu_h)/z(\bar\mu_h)$, owing to the very flat profile of data 
within errors, as it is seen from fig.~\ref{fig:fratiomuh} (right).
\item[-] \underline{Pole mass}: as the pole mass is affected by renormalon ambiguities, in the analysis based on the ratio method we compare the results obtained by using the NLO definition of the pole mass to the results found with the LO definition. Within the interpolation method, instead, we have also considered the alternative definition of $\Phi_{h \ell (s)}$ in terms of the pole mass (rather than the meson mass), again using either the NLO or the LO definition of the pole mass. In both cases, the differences are found to be small, at the level of 1 MeV, for the decay constants, as the sensitivity to the pole mass definition, which appears in the intermediate steps of the calculation, largely cancels out in the final determinations. The results for the ratio $f_{Bs}/f_B$ are practically unaffected.
\item[-] \underline{Mismatch of the strange quark mass in the static simulation}: as discussed in section 4.2, the static-strange correlators have been calculated with a value of the strange quark mass that was estimated from an analysis at fixed lattice spacing, and turned out to be larger with respect to the continuum limit estimate by approximately 22\% at $\beta=3.9$ and 13\% at $\beta=4.05$. In order to evaluate the systematic uncertainty due to this mismatch, we have analyzed the relativistic data for $\Phi_{h s}$ which are available at several values of the strange quark mass. By using the continuum estimate of the strange quark mass, $\Phi_{Bs}$ decreases by approximately 2\%. A similar effect can be thus expected for the static data. We have thus repeated the interpolation to the $b$ quark mass using for the static points results smaller by 2\%. We find that the $B_s$ decay constant decreases by 3$\mev$ and $f_{Bs}/f_B$ by $0.015$. We conservatively ignore the sign of the variation and consider these changes as a symmetric contribution to the systematic uncertainty. This uncertainty does not affect the ratio method, since in this case the static limit of $z_s$ and $z_s/z$ is exactly known.
\end{itemize}

For both methods we add in quadrature the systematic uncertainties and, finally, as shown in the last row of Table~\ref{tab:results}, we average the results of the two methods obtaining
\be
f_{Bs}=232(10)\,\mev\,, \qquad  \frac{f_{Bs}}{f_B}=1.19(5)\,,
\label{eq:risratio}
\ee
and for the $B$ decay constant, which is determined for each analysis as $f_B=f_{Bs}/(f_{Bs}/f_B)$,
\be
\qquad  f_B=195(12)\,\mev\,.
\label{eq:risfB}
\ee
These values are in agreement and improve the results obtained in~\cite{Blossier:2009hg} and \cite{Blossier:2009gd}.

As a byproduct of the analysis we also obtain the decay constants for the $D$ and $D_s$ mesons. In order to determine these quantities, we only consider three values for the heavy quark reference masses around the physical charm quark mass.
By interpolating to the physical value $\overline{m}_c(\overline{m}_c)=1.28(4)\gev$, obtained in~\cite{Blossier:2009gd}, we find
\be
f_D=212(8)\,\mev\,, \qquad \qquad f_{Ds}=248(6)\,\mev\,, \qquad \qquad \frac{f_{Ds}}{f_D}=1.17(5)\,,
\label{fDfDs}
\ee
to be compared with the results $f_D = 197(9)$ MeV, $f_{Ds} = 244(8)$ MeV and $f_{Ds}/f_D = 1.24(3)$ of~\cite{Blossier:2009bx}.
With respect to~\cite{Blossier:2009bx}, the present analysis is improved essentially for the reasons discussed for $f_B$ and $f_{Bs}$, namely: the statistics is increased for some ensembles, data at the finest lattice spacing ($\beta=4.2$) are now included, the continuum extrapolation is performed at fixed (reference) heavy quark masses. 
Moreover, as discussed for $f_{Bs}/f_B$, we perform the chiral extrapolation of $f_{h s}/f_{h \ell}$ either following HMChPT or a linear dependence on $\mu_\ell$. In ~\cite{Blossier:2009bx} the value $f_{Ds}/f_D= 1.24(3)$ was obtained from the HMChPT fit only, while the result given in eq.~(\ref{fDfDs}) is an average of $f_{Ds}/f_D= 1.21(2)$ from HMChPT and $f_{Ds}/f_D= 1.12(2)$ from the linear fit.
By considering both results we have increased the uncertainty associated to the chiral extrapolation.

\section{Conclusions}
We have presented a lattice determination of the $b$ quark mass and of the $B$ and $B_s$ decay constants, obtained with $N_f=2$ twisted mass Wilson fermions. Two methods have been employed, following and improving our previous analyses in~\cite{Blossier:2009hg} and~\cite{Blossier:2009gd}.

The first method is based on suitable ratios with exactly known static limit and smooth chiral and continuum limit. With respect to~\cite{Blossier:2009hg}, the present analysis includes data at four values for the lattice spacing, a larger statistics, and uses the published values for the quark mass renormalization constants~\cite{Constantinou:2010gr} and for the physical up/down and strange quark masses~\cite{Blossier:2010cr}.

The second method consists in interpolating between relativistic and static data. With respect to~\cite{Blossier:2009gd}, we added one ensemble at $\beta=4.2$, increased the statistics and, again, used the published values for the renormalization constants and light quark masses. A further improvement has been achieved by studying separately discretization effects and the (physical) dependence on the heavy quark mass. This has been done by performing the continuum extrapolation at fixed reference heavy quark mass.

The systematic uncertainties due to the chiral and continuum extrapolation and to the interpolation to the physical $b$ quark mass, as well as the sensitivity to the pole mass definition, have been carefully studied. An important uncertainty affecting the determination of the ratio $f_{Bs}/f_B$ and, in turn, of $f_B$, is introduced by the chiral extrapolation to the physical value of the average up/down quark mass. We note, in this respect, that given an assumption for the chiral extrapolation fitting function, i.e. either including or not the leading chiral logarithm, the results obtained  for the ratio $f_{Bs}/f_B$ by using the ratio and the interpolation method are in perfect agreement within each other (see Table~\ref{tab:results}). In order to reduce the uncertainty due to the chiral extrapolation, simulations at smaller values for the light quark masses, closer to their physical values, are needed.

The difference between the results obtained for $f_{Bs}$ by using the ratio and the interpolation method (approximately 5\%, see Table~\ref{tab:results}) provides an indication of the uncertainty due to the interpolation to the heavy $b$ quark mass. In this respect, the main advantage of the ratio method is that the static limit of the ratios is exactly known (by definition), so that the approach does not require a dedicated lattice simulation within the HQET.

The final results for the $b$ quark mass in the $\msb$ scheme and for the decay constants read
\bea 
&\mbar_b(\mbar_b)=4.29(14)\,\gev\,,&\nn\\
&&\nn\\
&f_B=195(12)\,\mev\,, \quad f_{Bs}=232(10)\,\mev\,\,, \quad \dfrac{f_{Bs}}{f_B}=1.19(5)\,.&
\eea
As a byproduct of the analysis we also obtain the results for the $f_D$ and $f_{Ds}$ decay constants
\be
f_D=212(8)\,\mev\,, \qquad \qquad f_{Ds}=248(6)\,\mev\,, \qquad \qquad \frac{f_{Ds}}{f_D}=1.17(5)\,,
\ee
which update and improve our previous determination~\cite{Blossier:2009bx}.

The only systematic uncertainty which is not accounted for by our results is the one stemming from the missing strange and
charm quark vacuum polarization effects. A comparison of our $N_f=2$ result for the $B$ and $B_s$ decay constants, to existing results from $N_f=2+1$ quark flavor simulations~\cite{Gamiz:2009ku,Albertus:2010nm} suggests that the error
due to the partial quenching of the strange quark is smaller at present than other systematic uncertainties. In this respect we mention that simulations with $N_f=2+1+1$ dynamical flavors are already being performed by ETMC and preliminary results for several flavor physics observables have been recently presented~\cite{Baron:2010bv,Baron:2010th}.

\vspace{0.3cm}
We thank all the ETMC members for fruitful discussions and the apeNEXT computer centers in Rome for their invaluable technical help. Some computation time has been used for that project on NW-Grid in Liverpool and HLRN in Berlin. G.H. acknowledges the support from the Spanish Ministry for Education and Science project FPA2009-09017, the Consolider-Ingenio 2010 Programme CPAN (CSD2007-00042), the Comunidad Aut\'onoma de Madrid
(HEPHACOS P-ESP-00346 and HEPHACOS S2009/ESP-1473) and the European project STRONGnet (PITN-GA-2009-238353). 
V.L., S.S. and C.T. thank MIUR (Italy) for partial financial support under the contracts PRIN08.
V.L. acknowledges the support of CNRS and the Laboratoire de Physique Th\'eorique d'Orsay, Universit\'e Paris-Sud 11, where part of this work was completed.
M.W. acknowledges support by the Emmy Noether Programme of the DFG (German
Research Foundation), grant WA 3000/1-1.

\section*{Appendix: Phenomenological analysis of the ratios $y$ and $z_s$} 
\vspace{0.4 cm}
The lattice results for the ratios $y$ and $z_s$ derived in section~3 deviate from the their static limit value, in the whole range of heavy quark masses from the charm mass value up to infinity, by only a small amount. Specifically, with the chosen value $\lambda \simeq 1.18$, the deviation is not more than 1.5\% and 4\% for $y$ and $z_s$ respectively.
By looking at the best fit curves of $y$ and $z_s$ as functions of $1/\bar\mu_h$, see figs.~\ref{fig:Mratiomuh} and~\ref{fig:fratiomuh} (left), one notices however a clear curvature, thus signalling a large $1/\bar\mu_h^2$ contribution in the heavy quark expansion compared to the linear term. In this appendix, we wish to show that this behavior is actually in good agreement with the predictions of the heavy quark expansion of $M_{h\ell}$ and $f_{hs}$ once one employs phenomenological or lattice based  estimates for the relevant coefficients.

We first discuss the phenomenological analysis of the ratio $y(\bar\mu_h)$ defined in eq.~(\ref{RATM}). By introducing in the expression for $y$ the heavy quark expansion for the heavy-light pseudoscalar meson mass,
\begin{equation}
M_{h\ell} = \mu_h^{\rm pole} + \bar{\Lambda} - \frac{(\lambda_1 + 3\lambda_2)}{2} \frac{1}{\mu_h^{\rm pole}} 
+ {\cal O}\left( \frac{1}{(\mu_h^{\rm pole})^2}\right)\,,
\label{Mhl-HQET}
\end{equation}
one finds
\begin{equation}
y = 1 - \bar{\Lambda} \frac{ \lambda^{\rm pole}-1 }{ \mu_h^{\rm pole} } + 
\left( \frac{(\lambda_1 + 3\lambda_2)}{2} (\lambda^{\rm pole}+1) + \bar{\Lambda}^2 \lambda^{\rm pole} \right) 
\frac{\lambda^{\rm pole}-1}{(\mu_h^{\rm pole})^2}\, ,
\label{y-form}
\end{equation}
where $\lambda^{\rm pole}$ is a smoothly varying function of $\bar\mu_h$ defined as $\lambda^{\rm pole} = \mu_h^{\rm pole}(\bar\mu_h)/ $ $\mu_h^{\rm pole}(\bar\mu_h/\lambda) = \lambda \, \rho(\bar\mu_h)/ \rho(\bar\mu_h/\lambda)$.

\begin{figure}[t]
\begin{center}
\includegraphics[width=0.80\textwidth]{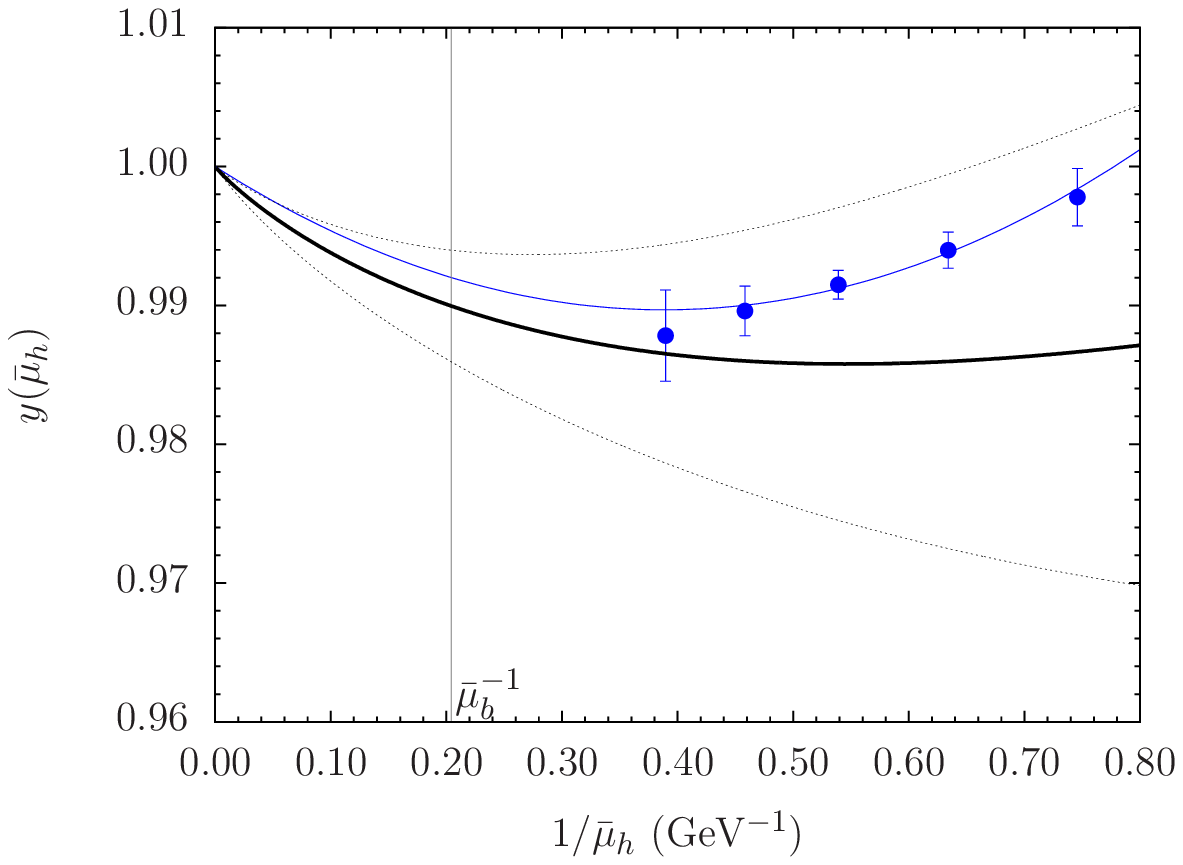}
\caption{\sl The lattice QCD results for $y$ (blue points) as a function of $1/\bar\mu_h$ are compared to the phenomenological estimate. For the latter, the black solid and dashed curves represent the mean value and the one-standard deviation band respectively.}
\label{fig:yPHEN}
\end{center}
\end{figure}

In order to estimate the ratio $y$ we considered the following phenomenological values for the HQET parameters
\begin{equation}
\bar{\Lambda} = 0.39(11) \; {\rm GeV} \, , \quad
\lambda_1 = -0.19(10) \; {\rm GeV}^2  \, , \quad
\lambda_2 = 0.12(2) \; {\rm GeV}^2  \, .
\label{PARAM-FOR-y}
\end{equation}
Notice that while $\lambda_2$ is rather precisely determined from the B-meson mass splitting $M_{B^*}^2 - M_B^2$, the values of $\bar{\Lambda}$ and $\lambda_1$ are inferred from the analysis of the inclusive semileptonic $B$-decays carried out in ref.~\cite{Gremm:1996yn}. Using eq.~(\ref{PARAM-FOR-y}) we obtain the phenomenological estimate of $y$ shown in fig.~\ref{fig:yPHEN}, which is compared to the lattice QCD results of section~3.
%%%%%%%%%%%%%%%%%%%%%%%%%%%%%%%%%%%%%%%%%%%%%%%%%%%%%
%%%%%%%%%%%%%%%%%%%%%%%%%%%%%%%%%%%%%%%%%%%%%%%%%%%%%

The above analysis shows that the lattice results for $y$ are consistent with the phenomenological estimate. In particular, one sees that the curvature of $y$ as a function of $1/\bar\mu_h$, observed in the lattice QCD data, is due to the fact that the coefficients of the linear and quadratic terms of the heavy quark expansion in eq.~(\ref{y-form}) are comparable in size and opposite in sign. As a result, the ratio $y$ has a minimum in the region around the charm quark mass, which is the one covered by the lattice data. 

A similar analysis can be also carried out for the ratio $z_s$, defined in eq.~(\ref{RATMZ}). In this case, the analysis is based on the heavy quark expansion for the pseudoscalar decay constant,
\begin{equation}
\Phi_{hs}(\bar\mu_h,\mu_b^*) = \frac{(f_{hs} \sqrt{M_{hs}})^{\rm QCD}}{C_A^{stat}(\bar\mu_h,\mu_b^*)} = 
\Phi_0(\mu_b^*) \, \left(1 + \frac{\Phi_1(\mu_b^*)}{\mu_h^{\rm pole}} +  
\frac{\Phi_2(\mu_b^*)}{(\mu_h^{\rm pole})^2} \right) + {\cal O}\left( \frac{1}{(\mu_h^{\rm pole})^3}\right) \, ,
\label{Phihs-HQET}
\end{equation}
where $\mu_b^*$ is the renormalization scale in the HQET. Using this expansion (and omitting for better clarity in the following the dependence on $\mu_b^*$), one finds
\begin{eqnarray}
y_s^{1/2} \, z_s & = &  \frac{ \Phi_{hs}(\bar\mu_h)}{ \Phi_{hs}(\bar\mu_h/\lambda)} = \nonumber \\
& = & 1 - \Phi_1 \frac{ \lambda^{\rm pole}-1 }{ \mu_h^{\rm pole} } - 
\left( \Phi_2 (\lambda^{\rm pole}+1) - \Phi_1^2 \lambda^{\rm pole} \right) 
\frac{\lambda^{\rm pole}-1}{(\mu_h^{\rm pole})^2}\, .
\label{zs-HQET}
\end{eqnarray}
The expansion for $z_s$ can be then obtained by combining the above expression with the heavy quark expansion for the ratio $y_s$. The latter has the same form of eq.~(\ref{y-form}) but with HQET parameters $\bar{\Lambda}_s$, $\lambda_{1s}$ and $\lambda_{2s}$ depending on the strange quark mass. For the purpose of the present exercise we take
\begin{equation} 
\bar{\Lambda}_s = \bar{\Lambda} + M_{B_s} - M_B \; , \quad
 \lambda_{1s} = \lambda_{1} \; , \quad \lambda_{2s} = \lambda_{2} \; .
\label{GUESS1}
\end{equation}

We are not aware of phenomenological estimates of the HQET parameters $\Phi_0$, $\Phi_1$ and $\Phi_2$. Therefore we consider in this case a set of values inferred from the lattice results for the heavy-light meson decay constants obtained by the HPQCD collaboration and presented in~\cite{DAVIES}. Their result for the static parameter is $\Phi_0 \simeq 0.60(4)~{\rm GeV}^{3/2} $, which is consistent with our calculation within lattice HQET, $\Phi_0 \simeq 0.67(4)~{\rm GeV}^{3/2} $, see eq.~(\ref{phistat}). The estimates for the parameters $\Phi_1$ and $\Phi_2$ can be derived, in turn, by requiring that eq.~(\ref{Phihs-HQET}) provides the HPQCD determinations for $f_{D_s}$ and $f_{B_s}$ at the physical charm and bottom quark masses, i.e. $f_{D_s} = 249(2)~{\rm MeV}$ and $f_{B_s} = 224(4)~{\rm MeV}$~\cite{DAVIES}. For this determination we also used the experimental values of the $D_s$ and $B_s$ meson masses~\cite{Nakamura:2010zzi} and the values of the charm and bottom quark masses obtained in section~3.1, namely $\bar\mu_c = 1.14(4)~{\rm GeV}$ and $\bar\mu_b = 4.91(15)~{\rm GeV}$. In conclusion we considered the values
\begin{equation}
\label{eq:phi_set1}
\Phi_0 = 0.60~{\rm GeV}^{3/2} \ , \quad
\Phi_1= - 0.48~{\rm GeV}  \ , \quad
\Phi_2=0.08~{\rm GeV^2} \, .
\end{equation}
%We also checked that varying the various inputs within the quoted errors does not lead to any significant change in the phenomenological estimates of $z_s$ below.

%%%%%%%%%%%%%%%%%%%%%%%%%%%%%%%%%%%%%%%%%%%%%%%%%%%%%
\begin{figure}[ht]
\begin{center}
\includegraphics[width=0.80\textwidth]{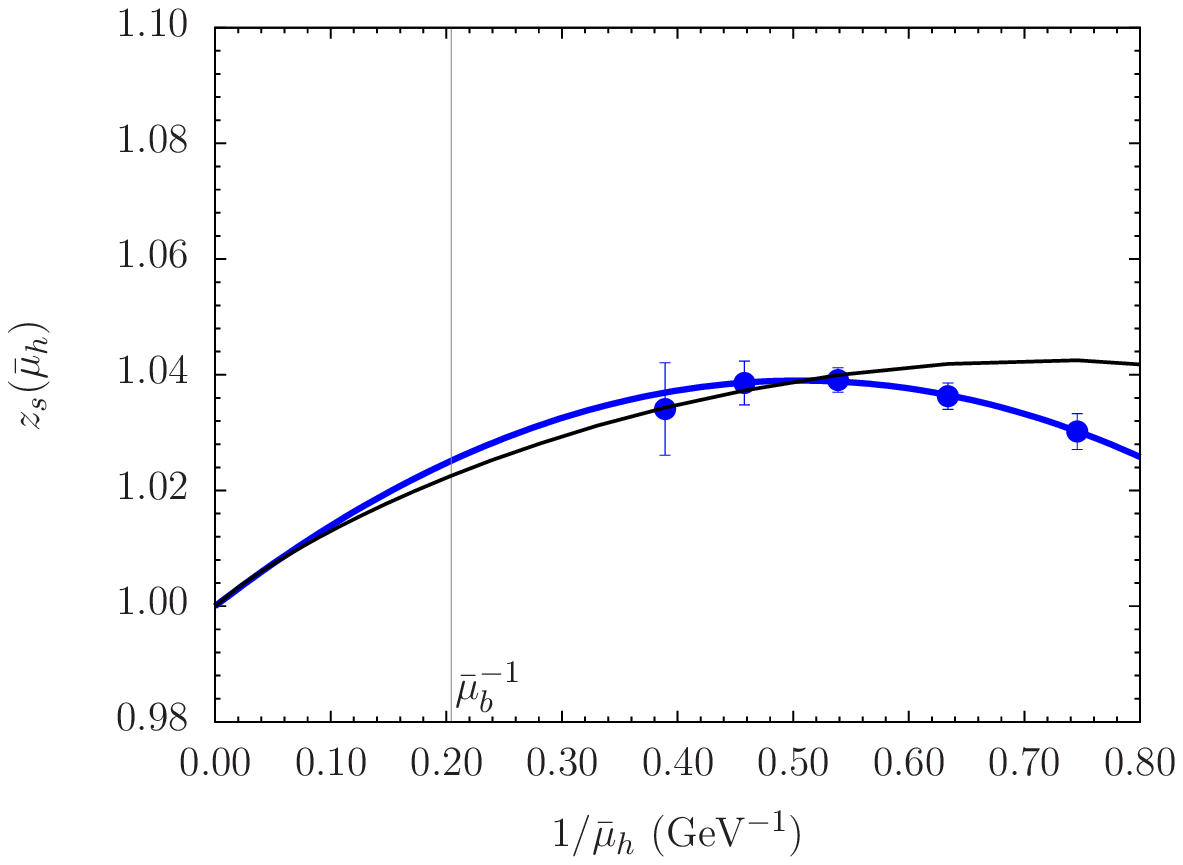}
\caption{\sl The lattice QCD results for $z_s$ (blue points) as a function of $1/\bar\mu_h$ are compared to the phenomenological estimate (black curve).} 
\label{fig:zsPHEN}
\end{center}
\end{figure}
%%%%%%%%%%%%%%%%%%%%%%%%%%%%%%%%%%%%%%%%%%%%%%%%%%%%%
The resulting phenomenological estimate of $z_s$ corresponding to the above set of values for the HQET parameters is shown in fig.~\ref{fig:zsPHEN} and compared to our determination from section~3. By also considering that a quantitative estimate of the uncertainties on the HQET parameters is beyond the scope of the present exercise we conclude, again, that the shape of the phenomenological curve, including its curvature, is well consistent with the lattice data.

\end{document}